# A dissipative quantum reservoir for microwave light using a mechanical oscillator

L. D. Tóth,[1,*] N. R. Bernier,[1,*] A. Nunnenkamp,[2] A. K. Feofanov,[1,†] and T. J. Kippenberg[1,‡]

[1] *Institute of Physics, École Polytechnique Fédérale de Lausanne (EPFL), CH-1015 Lausanne, Switzerland*

[2] *Cavendish Laboratory, University of Cambridge, Cambridge CB3 0HE, United Kingdom*

**Isolation of a system from its environment is often desirable, from precision measurements[1] to control of individual quantum systems; however, dissipation can also be a useful resource. Remarkably, engineered dissipation[2] enables the preparation of quantum states of atoms, ions or superconducting qubits[3–8] as well as their stabilization[9]. This is achieved by a suitably engineered coupling to a dissipative cold reservoir formed by electromagnetic modes. Similarly, in the field of cavity electro- and optomechanics[10], the control over mechanical oscillators utilizes the inherently cold, dissipative nature of the electromagnetic degree of freedom. Breaking from this paradigm, recent theoretical work[11–15] has considered the opposite regime in which the dissipation of the mechanical oscillator dominates and provides a cold dissipative reservoir to an electromagnetic mode. Here we realize this reversed dissipation regime[13] in a microwave cavity optomechanical system[16] and realize a quasi-instantaneous, cold reservoir for microwave light. Coupling to this reservoir enables to manipulate the susceptibility of the microwave cavity, corresponding to dynamical backaction control of the microwave field. Additionally, we observe the onset of parametric instability, i.e. the stimulated emission of microwaves (masing)[17,18]. Equally important, the reservoir can function as a useful quantum resource. We evidence this by employing the engineered cold reservoir to implement a large gain (above 40 dB) phase preserving microwave amplifier that operates 0.87 quanta above the limit of added noise imposed by quantum mechanics (×2 above the device's quantum limit). Beyond offering the manipulation of microwave fields, such a dissipative cold reservoir, when coupled to multiple cavity modes, forms the basis of microwave entanglement schemes[11], amplifiers with unlimited gain-bandwidth product[12] and the study of dissipative quantum phase transitions[19]. Moreover, combining such reservoir-mediated interaction with coherent dynamics allows for the realization of recently predicted non-reciprocal devices[15]. These devices could operate in the quantum regime when employing a cold reservoir as a quantum resource, thereby extending the available toolbox of quantum-limited microwave manipulation techniques[20–24].**

Reservoir engineering defies the notion that dissipation is necessarily detrimental for the utility of a quantum system. In fact, if carefully constructed, dissipation can relax the system of interest to a desired target quantum state, e.g. an entangled state. This pioneering insight was first theoretically conceived and studied in the context of trapped ions[2], experimentally first realized with trapped atomic ensembles[3] and later with trapped ions[4–6]. Moreover, reservoir engineering has recently also been realized in the context of circuit QED[7–9]. In these experiments the optical or microwave field provides a dissipative reservoir to the quantum systems. In cavity optomechanics[10], in which a mechanical oscillator and electromagnetic degree of freedom are parametrically coupled, analogous ideas have been developed and reservoir engineering for preparation of squeezed mechanical states has been theoretically proposed[25,26] and recently demonstrated[27–29]. As in the atomic physics case, the electromagnetic field acts as an engineered environment to the mechanical oscillator. In contrast, recent theoretical work[11–15] has considered the opposite scenario where the mechanical degree of freedom is employed to provide a dissipative, cold bath for light. This engineered bath can then be employed to achieve desirable quantum states of light or to modify the optical field properties. For example, such a dissipative reservoir for light can be exploited for amplification[12,13], entanglement generation[11] or dissipative squeezing of electromagnetic modes[14]. Moreover, it provides an ingredient to realize non-reciprocal devices[15] such as isolators, circulators or directional microwave amplifiers. For a sufficiently cold dissipative mechanical reservoir, non-reciprocal devices implemented in this manner can also operate in the quantum regime, i.e. with minimum added noise. Moreover, it is key to investigate non-equilibrium phase transitions in driven dissipative systems[19]. As a first step towards the full realization of such reservoir engineering-based schemes and devices, we realize here the reversed dissipation hierarchy by coupling an electromagnetic mode to a cold mechanical element whose energy decay rate exceeds that of the electromagnetic mode. We use this regime to manipulate the susceptibility of an electromagnetic mode and show the proof-of-principle of a microwave amplifier operated near the quantum limit for added noise.

In order to realize and study a cold, dissipative mechanical reservoir for an electromagnetic mode, we employ a microwave cavity optomechanical system[16]. We utilize a recently proposed scheme in which two microwave modes are coupled to the same mechanical oscillator[12,13]. One (auxiliary) electromagnetic mode is used to damp the oscillator via optomechanical sideband cooling[30,31] and engineer it into a cold bath for the other (primary) electromagnetic mode (Fig. 1A). A key ingre-dient to realize the reversed dissipation hierarchy in this manner is an optomechanical cooling rate of the auxiliary mode which greatly exceeds the electromagnetic decay rate of the primary microwave mode, necessitating vastly different decay rates of the employed microwave cavities. This is challenging to achieve with previously realized dual-mode circuits[29], since any parasitic coupling between the two modes opens a decay channel, equilibrating their decay in energy. Here, we address this challenge by engineering hybridized modes with inherently dissimilar decay rates arising from interference in the output channel (cf. Fig. 1B, C and SI). Reflecting the analogy to the bright and dark states of an atomic three-level $\Lambda$-type system, we subsequently refer to the device as an (electromechanical) $\Lambda$-type circuit.

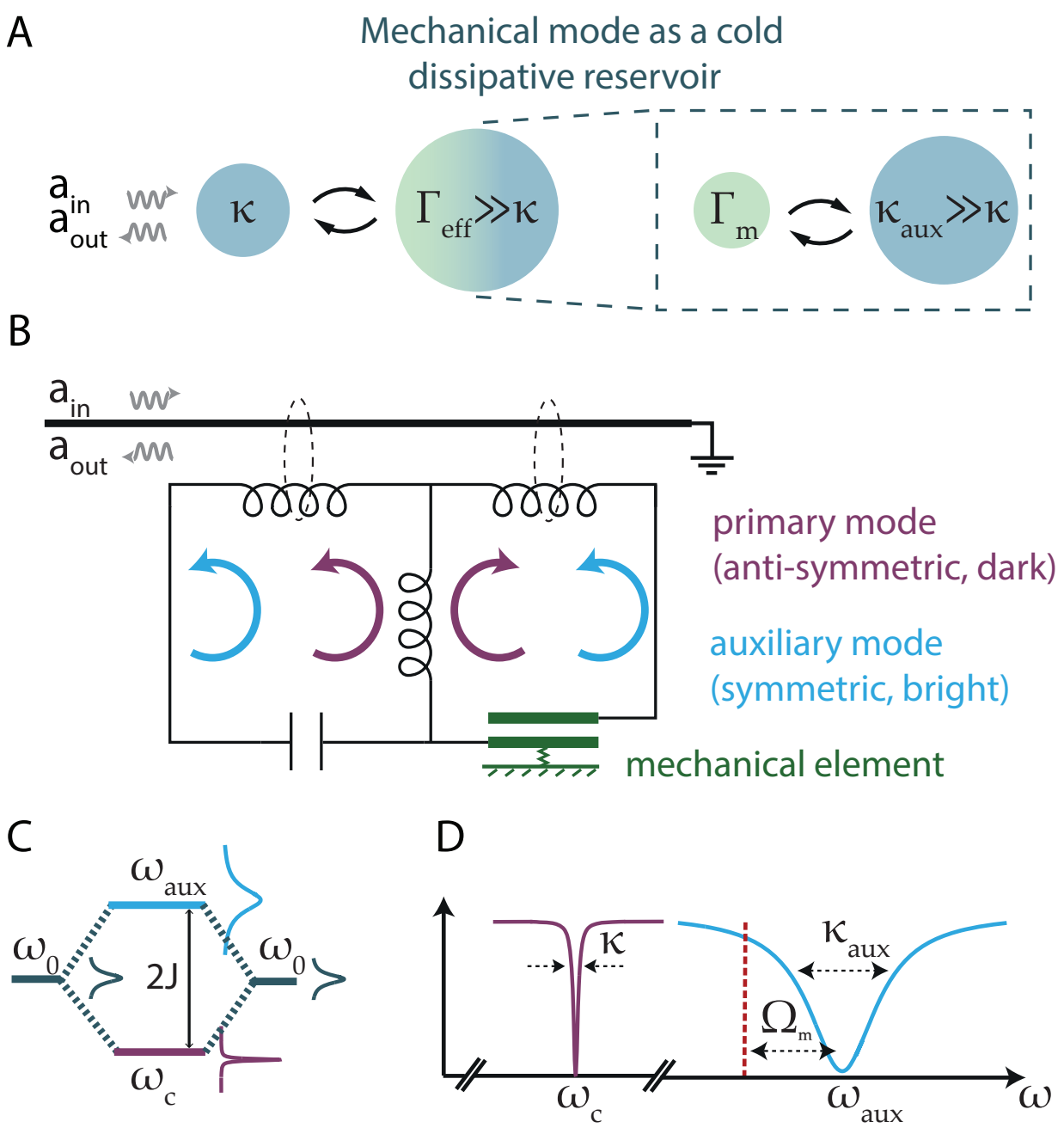

FIG. 1. **Realization of a cold, dissipative reservoir for microwave light in circuit optomechanics.** **A**. Schematic representation of a multi-mode electromechanical system in which a microwave mode with energy decay rate $\kappa$ is coupled to an engineered mechanical mode acting as a cold, dissipative reservoir with an effective energy decay rate $\Gamma_{\mathrm{eff}}$ much greater than $\kappa$. **B**. Circuit realization of the electromechanical system. Two lumped-element LC circuits – one containing a mechanically compliant capacitor – with matching resonance frequencies are inductively coupled and show normal-mode splitting, forming dark and bright modes (used as the primary and auxiliary modes, respectively) to achieve $\kappa_{\mathrm{aux}} \gg \kappa$. **C**. Visual representation of the mode structure and the resulting asymmetric dissipation rates, originating from the interference in the output coupling. **D**. Optomechanical sideband cooling the mechanical mode with the auxiliary, bright mode realizes a cold, dissipative mechanical reservoir for the primary, dark mode.

Specifically, we design an electromechanical $\Lambda$-type circuit using two LC resonators both coupled inductively to a common feedline, one of which has a mechanically compliant vacuum-gap capacitor[32] coupling mechanical vibrations to the microwave mode. The two resonators are strongly coupled through sharing a common inductor (cf. Fig. 1B). In terms of the bare modes, the resulting interaction Hamiltonian is given by

$$\hat{H}_{\mathrm{int}} = \hbar J(\hat{a}_1^\dagger \hat{a}_2 + \hat{a}_2^\dagger \hat{a}_1) + \hbar \tilde{g}_0 \hat{a}_1^\dagger \hat{a}_1 (\hat{b} + \hat{b}^\dagger), \quad (1)$$

where $\hat{a}_1$ and $\hat{a}_2$ designate the annihilation operators of the bare modes, $J$ the intermode coupling strength, and $\tilde{g}_0$ the vacuum electromechanical coupling strength to the first mode ($\hbar$ is the reduced Planck's constant). The symmetric and antisymmetric superpositions of the bare modes $\hat{a}_{\mathrm{s,a}} = \frac{1}{\sqrt{2}}(\hat{a}_1 \pm \hat{a}_2)$, diagonalize the intermode coupling $J$. In terms of these hybridized modes the interaction Hamiltonian is given by

$$\hat{H}_{\mathrm{int}} = \hbar J(\hat{a}_{\mathrm{s}}^\dagger \hat{a}_{\mathrm{s}} - \hat{a}_{\mathrm{a}}^\dagger \hat{a}_{\mathrm{a}}) + \hbar \frac{\tilde{g}_0}{2}(\hat{a}_{\mathrm{s}}^\dagger \hat{a}_{\mathrm{s}} + \hat{a}_{\mathrm{a}}^\dagger \hat{a}_{\mathrm{a}})(\hat{b} + \hat{b}^\dagger). \quad (2)$$

If the bare modes are degenerate, the eigenmodes have an energy difference of $2\hbar J$ (Fig. 1C) and are now, as a result of the interaction, both coupled to the micromechanical oscillator[33] with half the bare vacuum electromechanical coupling strength $\tilde{g}_0$. We consider here the limit $J \gg \Omega_{\mathrm{m}}$, implying that we can neglect the cross terms $\hat{a}_{\mathrm{s}}^\dagger \hat{a}_{\mathrm{a}}(\hat{b}+\hat{b}^\dagger)$ and other terms which are not resonant. Critically, the new eigenmodes will have dissimilar decay rates and form a bright (strongly coupled) and a dark (weakly coupled) mode resulting from interference of the bare-mode external coupling rates $\kappa_1^{\mathrm{ex}}$, $\kappa_2^{\mathrm{ex}}$ to the output channel (cf. SI). The symmetric, bright mode interferes constructively, leading to an external energy decay rate to the output feedline of $\kappa_{\mathrm{s}}^{\mathrm{ex}} = \kappa_1^{\mathrm{ex}} + \kappa_2^{\mathrm{ex}}$. Whereas, the antisymmetric, dark mode interferes destructively, leading to a decreased external coupling rate $\kappa_{\mathrm{a}}^{\mathrm{ex}} = |\kappa_1^{\mathrm{ex}} - \kappa_2^{\mathrm{ex}}|$. Physically, one can understand the difference by considering the topology of current flow in the modes (cf. Fig. 1B). The symmetric mode has current flowing in the same direction in both resonators, causing their external magnetic flux to create currents that add up, leading to an increased coupling rate to the feedline. The antisymmetric mode has current flowing in opposite directions in the two resonators, causing the external magnetic flux to create currents that cancel out, leading to a suppression in the external coupling to the feedline (cf. Fig. 1B, C and SI). For bare coupling rates similar in magnitude ($\kappa_1^{\mathrm{ex}} \approx \kappa_2^{\mathrm{ex}}$), this enforces the coupling-rate hierarchy $\kappa_{\mathrm{a}}^{\mathrm{ex}} \ll \kappa_{\mathrm{s}}^{\mathrm{ex}}$ necessary to achieve the reversed dissipation hierarchy with the present scheme[13]. In the remaining of the letter, we refer to the dark mode as the primary mode and the bright mode as the auxiliary mode with resonance frequencies $\omega_{\mathrm{c}}$ and $\omega_{\mathrm{aux}}$ and energy decay rates $\kappa$ and $\kappa_{\mathrm{aux}}$, respectively (Fig. 1).

We realize the electromechanical $\Lambda$-type circuit experimentally by fabricating two lumped-element LC circuits coupled to each other via a common inductor, made from thin-film aluminium on a sapphire substrate (cf. SI for the details of fabrication, design and full circuit parameters). The primary and auxiliary modes have resonance frequencies $(\omega_\mathrm{c}, \omega_\mathrm{aux}) = 2\pi \cdot (4.26, 5.48)$ GHz with total energy decay rates $(\kappa, \kappa_\mathrm{aux}) = 2\pi \cdot (118, 4478)$ kHz, respectively (thus, $\kappa_\mathrm{aux}/\kappa \approx 38$). This clear hierarchy in the energy relaxation rates indeed originates from the vastly dissimilar engineered external coupling to the feedline, with $\kappa_\mathrm{ex} = 2\pi \cdot 42$ kHz $\approx 1\%$ of $\kappa_\mathrm{aux}^\mathrm{ex}$. The mechanical resonator is a parallel-plate capacitor with a suspended top electrode, having a resonance frequency of the fundamental flexural mode $\Omega_\mathrm{m} = 2\pi \cdot 5.33$ MHz and a decay rate $\Gamma_\mathrm{m} = 2\pi \cdot 30$ Hz. This mechanical mode couples to both the primary and auxiliary modes with a vacuum electromechanical coupling strength $g_0 = \tilde{g}_0/2 = 2\pi \cdot 60$ Hz. Details of the calibration procedure are described in Supplementary Information. Importantly, the resolved sideband regime is still attained for both microwave modes, i.e. $\Omega_\mathrm{m} > \kappa_\mathrm{aux} \gg \kappa$. Fig. 2A, B show an optical image of the fabricated circuit and a scanning electron micrograph of the drum-type capacitor, respectively. The simplified measurement setup is shown in Fig. 2C. In brief, the device is mounted on the base plate of a dilution refrigerator and cooled to a base temperature of ca. 10 mK. The microwave input lines are heavily attenuated to suppress residual thermal noise and, in addition, filter cavities are employed to remove unwanted frequency noise from the applied tones (cf. SI). After amplification with a commercial high-electron-mobility transistor (HEMT) amplifier mounted on the 3 K plate, the signal is measured with a spectrum analyzer or a vector network analyzer. To realize the reversed dissipation regime (i.e. to prepare a cold, dissipative mechanical bath) we follow the approach outlined in Ref. 13 and use optomechanical sideband cooling[30,31] to prepare the mechanical oscillator as a strongly dissipative, cold reservoir. We proceed by pumping the auxiliary mode on the lower motional sideband (Fig. 1D). We strongly damp the mechanical oscillator to $\Gamma_\mathrm{eff} \approx 2\pi \cdot 500$ kHz (corresponding to a mean intra-cavity photon number of $n_\mathrm{c}^\mathrm{aux} \approx 1.5 \cdot 10^8$), while still remaining in the weak-coupling regime for the auxiliary mode. Thereby, we achieve the reversed dissipation hierarchy with respect to the primary, high-Q mode, since $\Gamma_\mathrm{eff} \gg \kappa$. The effective temperature of this reservoir and its utility as a quantum resource are studied below in the manuscript.

We first study the modified microwave cavity susceptibility resulting from the dissipative cold reservoir, i.e. the dynamical backaction on the microwave light. In the reversed dissipation hierarchy, the engineered bath provided by the mechanical resonator modifies the response of the electromagnetic mode when a microwave tone is applied[13]. For a pump detuned by $\Delta$ from the (primary) microwave cavity resonance, the frequency and the decay rate of the (primary) microwave mode shift by

$$\delta\omega_\mathrm{om} = \mathrm{Re}\Sigma \quad \text{and} \quad \kappa_\mathrm{om} = -2\mathrm{Im}\Sigma, \tag{3}$$

called mechanical spring effect and mechanical damping,

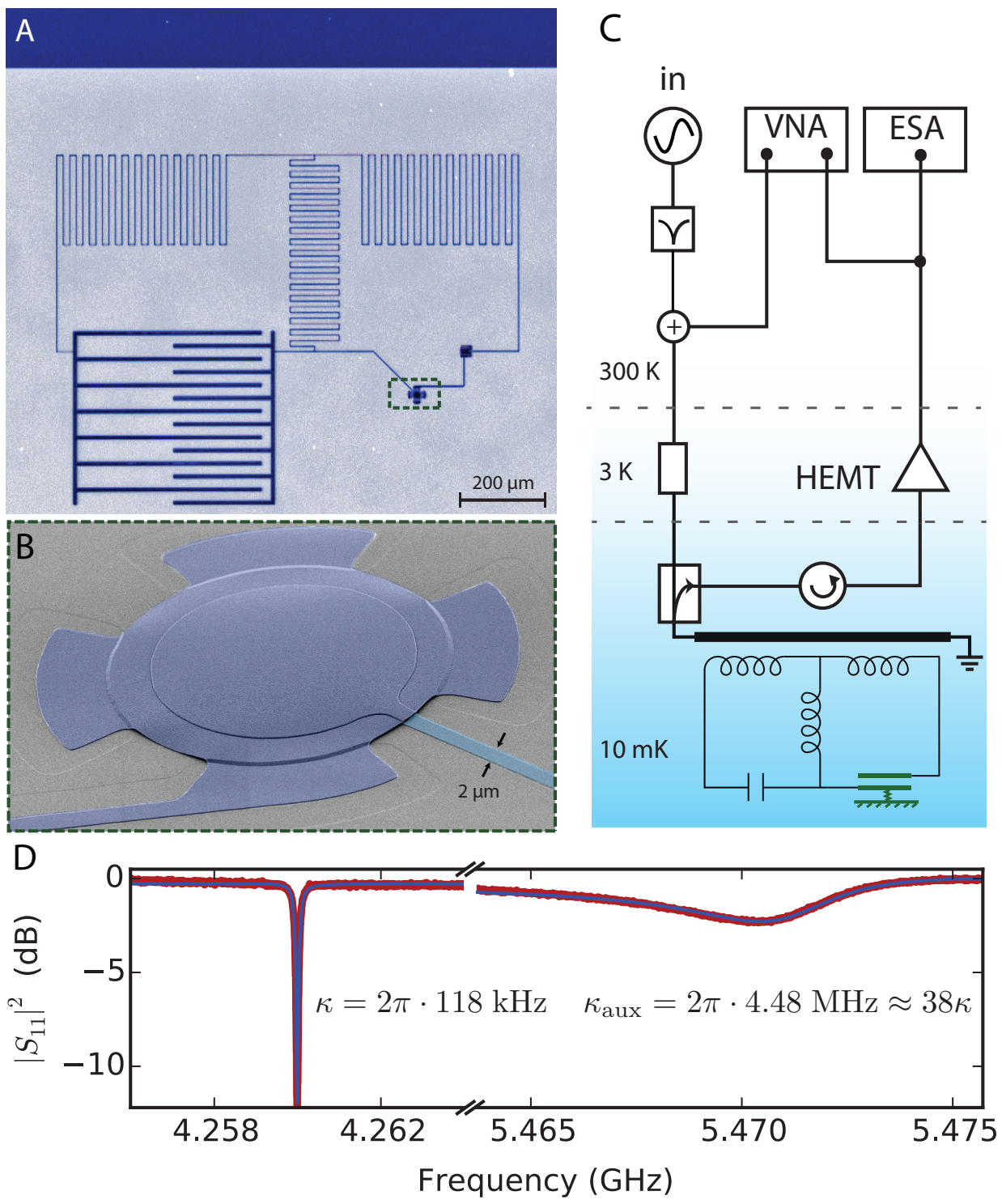

FIG. 2. **Device, experimental setup, and characterization of the electromechanical Λ-type circuit for achieving the reversed dissipation hierarchy. A**. Inverted-colour optical micrograph of the circuit consisting of two coupled LC resonators, one having a mechanically compliant capacitor. Blue regions show aluminium and grey regions that are the exposed sapphire substrate. **B**. False-colour scanning electron micrograph of the mechanically compliant drum capacitor. **C**. Simplified schematics of the measurement setup with the circuit. The (multiple) input lines are filtered and attenuated at various stages before reaching the device mounted in a dilution refrigerator. Both the coherent and the spectral response can be measured. **D**. Linear response measurement of the device revealing the symmetric (bright, used as the auxiliary) and anti-symmetric (dark, used as the primary) microwave modes.

respectively. The self-energy $\Sigma$ is defined as

$$\Sigma = -ig^2 \left( \frac{1}{\Gamma_\mathrm{eff}/2 + i(\Delta + \Omega_\mathrm{m})} - \frac{1}{\Gamma_\mathrm{eff}/2 + i(\Delta - \Omega_\mathrm{m})} \right), \tag{4}$$

where $g = g_0\sqrt{n_\mathrm{c}}$ is the effective electromechanical coupling rate enhanced by the mean intracavity photon number of the primary mode $n_\mathrm{c}$. This effect can be viewed as radiation pressure dynamical backaction[17,31,37] onto the *microwave* mode. This leads to a change in the reflection from the microwave cavity, due to a modification of its susceptibility (defined as $\hat{a}_\mathrm{out}(\omega) = S_{11}(\omega)\hat{a}_\mathrm{in}(\omega)$, where $\hat{a}_\mathrm{in,out}(\omega)$ are Fourier domain operators associated with the input and output fields, cf. Fig. 1B). The susceptibil-

ity becomes

$$S_{11}(\omega) = \frac{\kappa_0 + \kappa_{\rm om} - \kappa_{\rm ex} - i2(\omega - \omega_{\rm c}')}{\kappa_0 + \kappa_{\rm om} + \kappa_{\rm ex} - i2(\omega - \omega_{\rm c}')}, \quad (5)$$

with the modified resonance frequency $\omega_{\rm c}' = \omega_{\rm c} + \delta\omega_{\rm om}$.

The engineered reservoir therefore supplies a way to tailor the susceptibility of the primary electromagnetic mode, which we can directly probe using a coherent response measurement. First, we fix the detuning to either motional sideband of the primary mode, and measure $S_{11}(\omega)$ while the power is varied. For this choice of detuning (i.e. $\Delta = \mp\Omega_{\rm m}$), we have $\delta\omega_{\rm om} = 0$ (neglecting the term $\propto \Gamma_{\rm eff}^2/\Omega_{\rm m}$) and the change in the microwave decay rate simplifies to

$$\kappa_{\rm om} = \pm\mathcal{C}\kappa, \quad (6)$$

directly proportional to the cooperativity $\mathcal{C} = \frac{4g^2}{\kappa\Gamma_{\rm eff}}$. Fig. 3A, B show the linear response for a tone on the lower and upper sideband for various pump powers. The width of the resonance, corresponding to the cavity decay rate, increases (for $\Delta = -\Omega_{\rm m}$) or decreases (for $\Delta = +\Omega_{\rm m}$) linearly with $\mathcal{C}$, accordingly. Strikingly, the depth of reflection on resonance $|S_{11}(\omega_{\rm c})|^2$ varies significantly to reflect this change (Fig. 3C). The effective *internal* loss of the cavity $\kappa_0+\kappa_{\rm om}$ can be tuned on demand by changing the coupling to the dissipative reservoir via the pump tone. While the microwave cavity is initially undercoupled ($\kappa_{\rm ex} < \kappa_0$), pumping on the upper sideband reduces the effective internal loss and increases the depth on resonance until the cavity becomes critically coupled (the effective internal loss matches the external coupling, i.e. $\kappa_0 + \kappa_{\rm om} = \kappa_{\rm ex}$). Increasing the power further, the cavity becomes overcoupled ($\kappa_0 + \kappa_{\rm om} < \kappa_{\rm ex}$) and resonant reflection increases again. When $\kappa_0 + \kappa_{\rm om}$ becomes negative, there is net internal gain: the absorptive feature in the cavity reflection becomes a peak, indicating amplification of the reflected microwave signal. By pumping on the lower sideband ($\Delta = -\Omega_{\rm m}$), extra damping is introduced and the resonance becomes increasingly undercoupled. The mechanical mode provides a cold, dissipative bath for the microwave resonator, down-converting the cavity photons to the pump. In Fig. 3C we plot the resonant reflection and observe good agreement with the expected dependence according to equation (5). For the data corresponding to the pump tuned to the lower motional sideband ($\Delta = -\Omega_{\rm m}$), the depth of the resonance is systematically lower than expected, due to a decrease in the intrinsic microwave cavity loss in the presence of a strong pump[34]. In Fig. 3D, E, we keep the pump power constant and sweep the detuning $\Delta$, to measure the mechanical spring and damping effects. For the frequency shift $\delta\omega_{\rm om}$, intrinsic non-linearities redshift the resonance frequency in an asymmetric fashion, providing a different background for the red and blue sidebands. The spring effect agrees well with the prediction from equation (3) when the two sidebands are fitted independently with different constant offsets. We note that the mechanical spring effect as a function of detuning has the opposite parity compared to the better known case of the optical spring effect[10].

In the remainder of the letter, we demonstrate the cold nature of the dissipative mechanical reservoir by studying the noise properties of the system. To this end, we fix the microwave drive to the upper sideband ($\Delta = +\Omega_{\rm m}$) and study the regime where the pump introduces net gain in the microwave cavity ($\kappa_0 + \kappa_{\rm om} < 0$). We use a different (second) device for this analysis, with optimized properties, due to higher coupling strength ($g_0 = 2\pi \cdot (106, 79)$ Hz for the primary and auxiliary modes, respectively) and the primary mode being overcoupled ($\kappa_{\rm ex}/\kappa = 0.76$), cf. SI. In Fig. 4A, the emitted noise spectra of the microwave cavity are shown for different pump powers. The measured power spectrum is rescaled to the symmetrized cavity output field spectrum[38] $\bar{S}_{aa}(\omega)$ in units of photons per second (i.e. flux) per unit bandwidth, using the noise temperature of the HEMT as an absolute noise reference (cf. SI). As the pump compensates for the losses, the width of the emitted noise spectrum, corresponding to the cavity linewidth $\kappa_{\rm eff} = (1 - \mathcal{C})\kappa$, decreases linearly with the pump power towards zero (at unity cooperativity $\mathcal{C} = 1$), cf. inset of Fig. 4B. In this below-threshold regime, the peak photon flux spectral density emitted from the cavity increases with power, as the vacuum noise and the residual thermal microwave noise (consisting in both a finite residual occupancy $n_{\rm eff}$ of the dissipative mechanical reservoir and a finite thermal microwave occupation of the cavity $n_{\rm cav}$) are amplified according to

$$\bar{S}_{aa}(\omega) = \kappa_{\rm ex}\frac{(\kappa_{\rm ex} - \kappa_{\rm eff})\frac{1}{2} + \mathcal{C}\kappa(n_{\rm eff} + \frac{1}{2}) + \kappa_0(n_{\rm cav} + \frac{1}{2})}{(\frac{\kappa_{\rm eff}}{2})^2 + (\omega - \omega_{\rm c})^2} \quad (7)$$

where the thermal input noise is neglected and only the amplified noise is considered. We analyze the noise properties of the device in detail below when considering amplification and added noise. We find the residual thermal occupation of the dissipative reservoir to be $n_{\rm eff} = 0.66$, when neglecting $n_{\rm cav}$. Equation (7) then implies that 60% of the emitted noise from the cavity is amplified vacuum fluctuations, when $\mathcal{C} \to 1$.

For $\mathcal{C} = 1$ and greater pump powers, the microwave mode undergoes self-sustained oscillations. This regime leads to a parametric instability and electromechanical maser action of the microwave mode, via stimulated emission of microwave photons into the microwave cavity. The salient features of maser action are a transition from sub- to above-threshold masing, as well as linewidth narrowing. These observations are analogues to the radiation-pressure-induced parametric instability of a mechanical mode in the normal dissipation regime[17,18,39] (phonon lasing[40]). In the experiments a clear threshold behavior, characteristic of masing, is demonstrated when the emitted noise abruptly increases in strength at $\mathcal{C} = 1$ (cf. Fig. 4B). Such microwave lasing in superconducting circuits has previously been demonstrated using a single artificial atom[41]. Due to the strong photon population generated by masing, nonlinearities of the cavities red-shift the frequency of emission. This clearly distinguishes masing from the mechanical parametric instability (i.e. phonon lasing[40]) in the normal dissipation hierarchy, as in the latter case the emission does not follow the cavity but has a constant detuning of $-\Omega_m$ with respect to the pump.

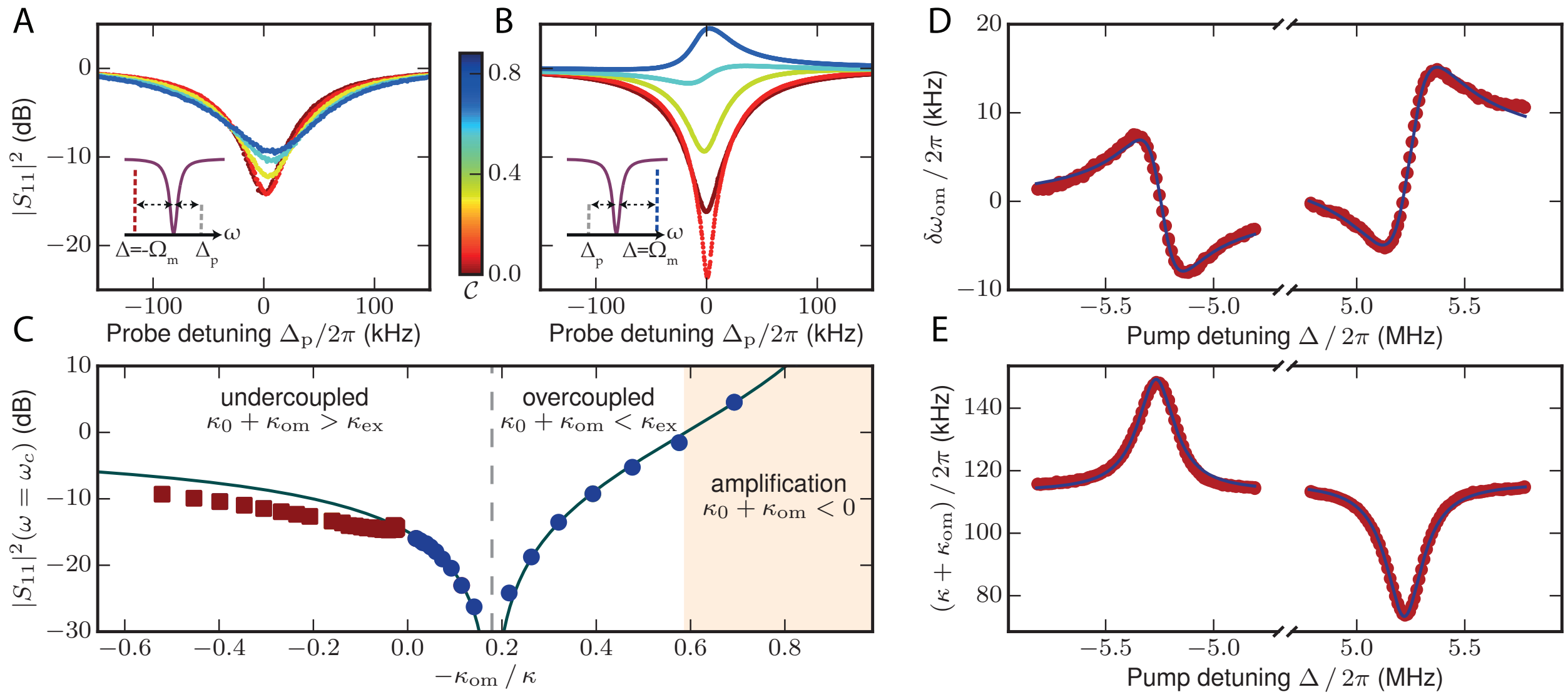

FIG. 3. **Manipulation of the microwave susceptibility with an engineered mechanical reservoir. A**, **B**. The modification of the linear response of the microwave cavity when a pump tone is placed on the lower and upper motional sideband of the primary mode (i.e. $\Delta = \mp\Omega_m$, panel A and B, respectively, cf. insets), shown for various values of the multiphoton cooperativity ($\mathcal{C} = 1$ corresponds to a mean intracavity photon number of $n_c \approx 5\cdot 10^6$). The slight shift of the peaks comes from residual spring effect ($\delta\omega_{om} \neq 0$). **C**. The depth of the resonance changes depending on the effective internal losses $\kappa_0 + \kappa_{om}$. Measurements with a pump on the lower and upper sideband ($\Delta = -\Omega_m$ for the red squares and $\Delta = +\Omega_m$ for the blue circles), and a theoretical fit are shown. The cavity is originally undercoupled. When pumped on the upper motional sideband, it first becomes critically coupled, then overcoupled as the power of the pump tone is increased. For $\kappa_0 + \kappa_{om} < 0$, there is a net gain and the electromechanical system acts as a phase-preserving microwave amplifier. **D**, **E**. Using a fixed pump power, the detuning $\Delta$ of the pump tone is swept and the change in the microwave resonance frequency and decay rate is recorded. The theoretical fit corresponds to equation (3), showing good agreement with the experimental data.

Below the masing threshold, the microwave mode coupled to the dissipative bath acts as a phase-insensitive parametric amplifier[13,42] for incoming signals. For $\kappa_0 + \kappa_{om} < 0$, there is a net internal gain and the susceptibility $S_{11}(\omega)$ develops a peak, implying that reflection is larger than input for signals within the resonance bandwidth (in-band). The power gain of the amplifier is defined as the resonance peak height above the background, given by (cf. SI)

$$\mathcal{G}(\omega_c) = |S_{11}(\omega_c)|^2 = \left|\frac{\left(2\frac{\kappa_{ex}}{\kappa} - 1\right) + \mathcal{C}}{1-\mathcal{C}}\right|^2. \quad (8)$$

The bandwidth of the amplifier is the linewidth of the microwave resonance, given by $\kappa_{eff} = (1-\mathcal{C})\kappa$. In order to measure the gain, bandwidth and noise properties of the amplifier, we inject, in addition to the pump tone on the upper sideband ($\Delta = +\Omega_m$), a weak signal tone (swept around the cavity resonance) and measure the reflected signal as a function of the pump tone power. With increasing pump power, a narrowing of the cavity bandwidth (Fig. 5A) is observed, as well as an increase in the power of the reflected signal (i.e. gain). By fitting the reflected power as a function of detuning, the gain and bandwidth as a function of cooperativity are extracted, and found to be in good agreement with the theoretical predictions given by equation (8) (Fig. 5B). The observed gain exceeds 42 dB. Next, we study the added noise of the dissipative amplification process. The added noise $\mathcal{N}$ is given by the noise output of equation (7) without the input noise and divided by the gain $\mathcal{G}(\omega_c)$. On resonance, it is found to be (cf. SI)

$$\mathcal{N}(\omega_c) = \frac{4\mathcal{C}\frac{\kappa_{ex}}{\kappa}\left(n_{eff}+\frac{1}{2}\right) + 4\frac{\kappa_{ex}\kappa_0}{\kappa^2}\left(n_{cav}+\frac{1}{2}\right)}{\left(\mathcal{C}-1+\frac{2\kappa_{ex}}{\kappa}\right)^2}, \quad (9)$$

which, in the high gain limit ($\mathcal{C} \to 1$), simplifies to $\mathcal{N}(\omega_c) \to \frac{\kappa_0}{\kappa_{ex}}(n_{cav}+\frac{1}{2}) + \frac{\kappa}{\kappa_{ex}}(n_{eff}+\frac{1}{2})$. This quantity can be measured by recording the improvement of the signal-to-noise ratio (SNR) of amplification in and out of the bandwidth of our device. This directly compares the noise performance of our device with the commercial HEMT amplifier, which is used as a calibrated noise source (the noise temperature of the HEMT is measured separately at $\omega_c$ and found to be $3.95 \pm 0.02$ K, corre-sponding to $20.0 \pm 0.1$ quanta, cf. SI). In Fig. 5 C, the gain of the device is compared to the noise output of the chain, normalized to the HEMT noise background. This calibration was corroborated by a second, *independent* calibration technique, which uses the scattered power in the motional sideband in conjunction with the knowledge of the intracavity photon number and $g_0$ (cf. SI). The relative gain of the signal exceeds the relative noise by over 12 dB. From this apparent SNR improvement, one must subtract the insertion loss of the components between the device and the HEMT, measured independently at 77 K to be 1.6 dB (cf. SI). The analysis reveals therefore that the optomechanical amplifier provides more than 10 dB of improvement over the SNR of the HEMT. The inferred added noise on resonance is shown as a function of gain in Fig. 5 D; in the high-gain limit, it is a constant value of $\mathcal{N}(\omega_c) = 1.68 \pm 0.02$ quanta per second per unit bandwidth (with the uncertainty given by statistical fluctuations). Using equation (9) and assuming $n_{\rm cav} = 0$, the effective occupation of the dissipative reservoir is found to be $n_{\rm eff} = 0.66$. However, the strong cooling pump increases the temperature of the cavity thermal bath to an occupation $n_{\rm cav} = 1.03$, obtained from measuring the emitted thermal noise of the microwave cavity. (cf. SI). Taking the residual cavity thermal noise into account, the estimate for the mechanical occupation is reduced to $n_{\rm eff} = 0.41$. This demonstrates that the dissipative mechanical reservoir constitutes a quantum resource. We note that even in the case when all the thermal noise sources are reduced to zero (i.e. $n_{\rm eff} = n_{\rm cav} = 0$), the added noise of the amplifier is

$$n_{\rm DQL} = \frac{1}{2} + \frac{\kappa_0}{\kappa_{\rm ex}}, \qquad (10)$$

which, we call the device quantum limit and deviates from 1/2 due to the finite internal dissipation rate $\kappa_0$. For the present system the device quantum limit amounts to 0.81 quanta for the coupling ratio of $\kappa_{\rm ex}/\kappa = 0.76$, which is only 0.87 quanta below the added noise we measure. Compared to other electromechanical amplifiers[43], operating in the normal dissipation regime, the preparation of an engineered cold, dissipative mechanical bath enables to overcome[13] the large added noise while increasing the bandwidth of amplification. It is interesting to compare the present amplifier scheme, relying on a dissipative reservoir, to the microwave parametric amplifiers as used in circuit QED. In the latter case, typically both idler and signal are resonant with one or more microwave cavities[20–22]. As gain increases, this leads to a simultaneous increase in both the signal and idler mode population. In contrast, while the present amplifier scheme uses a parametric interaction as well, the large dissipation rate for the (mechanical) idler mode only leads to the generation of a signal photon (microwave field), suppressing the idler, a situation akin to a Raman-type interaction found in nonlinear optics[44].

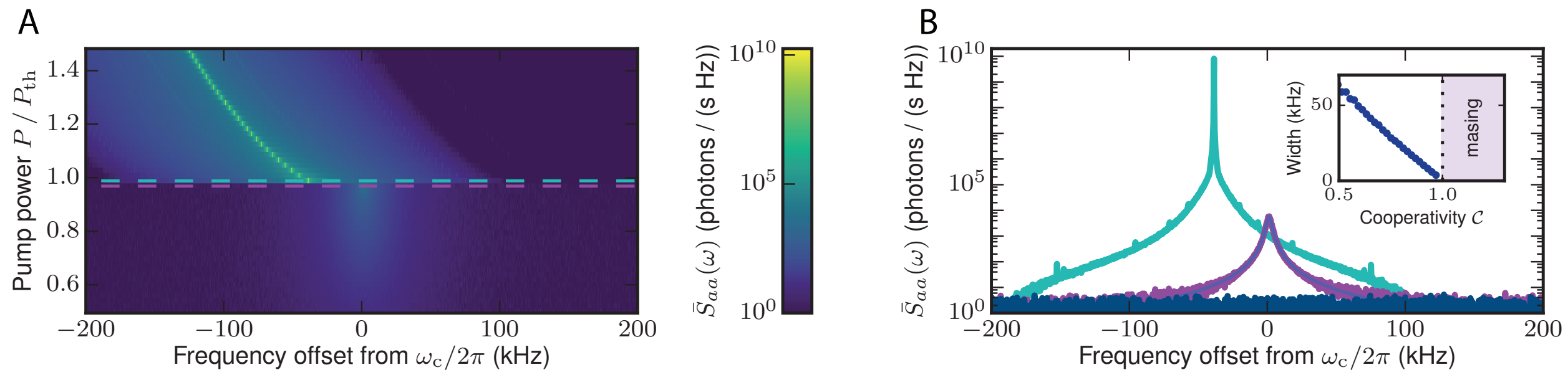

FIG. 4. **Amplified vacuum fluctuations and parametric instability of the microwave mode (masing). A**. Noise spectrum of the cavity emission as a function of the power of a pump on the upper motional sideband. The spectrum is rescaled using the HEMT amplifier as a reference (cf. main text). Above a certain threshold power, the microwave mode undergoes self-sustained oscillations, characteristic to masing. The vertical axis is normalized by the pump power at masing threshold $P_{\rm th}$, equivalent to the cooperativity $\mathcal{C} = P/P_{\rm th}$ below threshold. **B**. Two examples of emission from the microwave mode at the input of the HEMT, below and above masing threshold (line cuts of A), as well as a reference measurement of the background without the pump (dark blue). The inset shows the emission narrowing below threshold. In this regime, the noise emission is composed of amplified vacuum and thermal fluctuations, described by equation (7). The analysis reveals that the measured noise is composed of 60% amplified vacuum noise below threshold.

In summary, we have implemented and studied a new regime of circuit electromechanics by coupling an electromagnetic cavity mode to an engineered cold dissipative reservoir formed by a mechanical oscillator. This allows the control of the internal losses of the cavity and the observation of backaction-induced amplification, deamplification, and masing of the microwave field.

The observation of near-quantum-limited *dissipative* amplification extends the available quantum information manipulation toolkit realized using Josephson junctions[20–22,24]. An advantage of the mechanical nonlinearity is the large dynamic range over which amplification can be achieved. While the present amplifier is not tunable, recent advances in circuit electromechanics have demonstrated tunable vacuum-gap capacitors that would provide such functionality[48]. The observed reservoir-mediated microwave damping process may provide a way to remove residual thermal occupancy from the microwave cavity, akin to cooling schemes developed in superconducting circuits[49]. Moreover, this control over internal dissipation enables all-electromechanical tuning of the coupling of the microwave resonator to the feedline, offering the potential for an electromechanical reconfigurable network[50]. While the present schemes employed only one pump tone, dual tone pumping would lead to the dissipative preparation of squeezed states of the microwave field[14]. Viewed more broadly, the realization of cold, dissipative mechanical reservoir provides a central ingredient for novel electromechanical devices. Indeed, the circuit can be extended to multiple ($> 2$) microwave resonators mutually coupled to one or more mechanical oscillators[33], enabling to realize dissipative cavity-cavity interactions that are at the heart of recent schemes to create microwave entangled photon fields[14] or amplifiers with infinite gain-bandwidth product[12]. Combining simultaneously dissipative and coherent interactions in one single circuit has recently been predicted to allow the realization of a set of quantum-limited non-reciprocal devices in the microwave domain, such as isolators and directional amplifiers[15]. Such non-reciprocal devices can be of use for the rapidly expanding field of circuit QED[51,52].

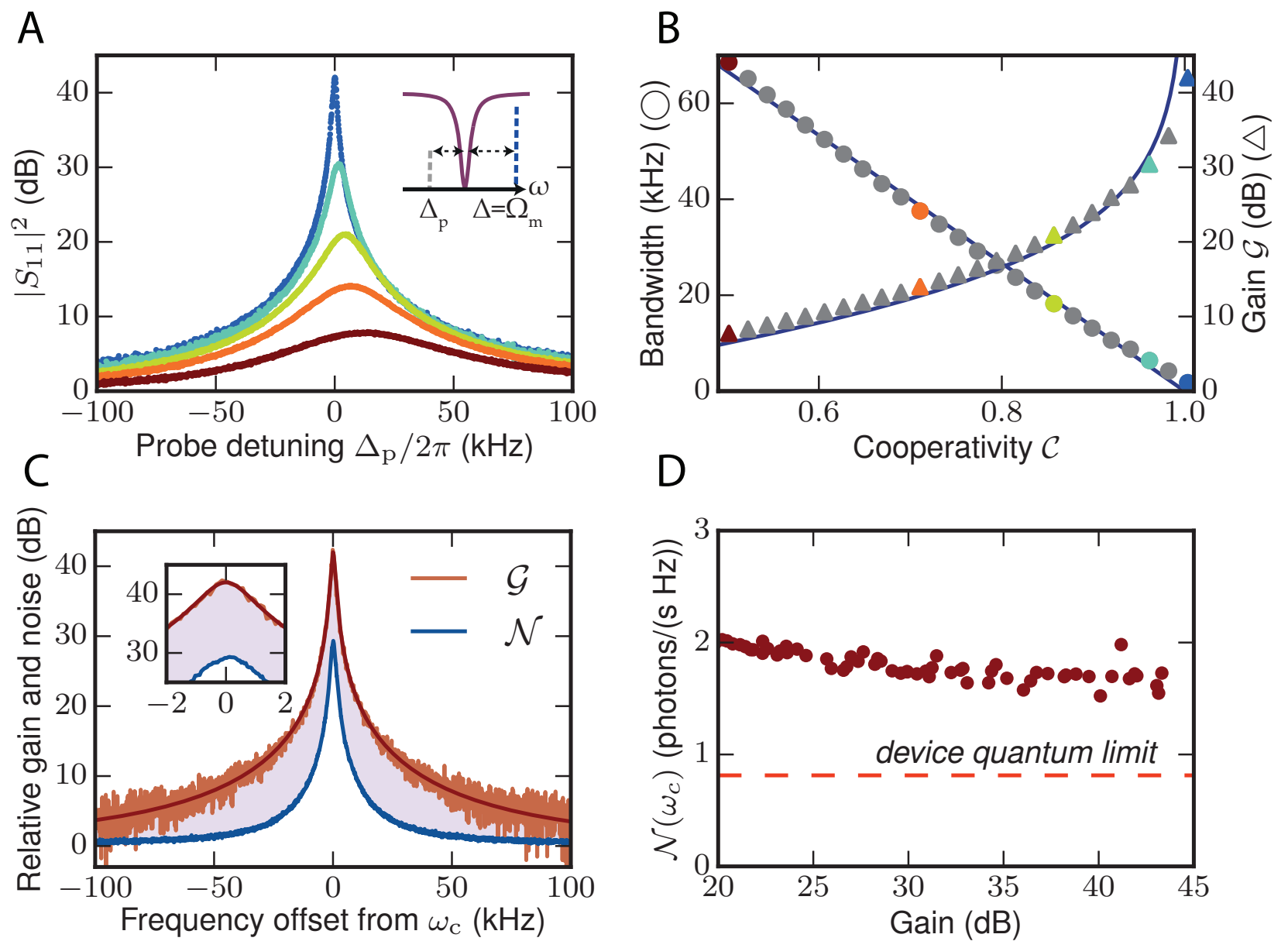

FIG. 5. **Near-quantum-limited phase-preserving amplification.** **A**. Linear response of the cavity, with increasing powers of the pump on the upper sideband from red to blue. **B**. Power gain (triangles) and bandwidth (circles) of the amplifier extracted from a fit of the linear response, as a function of the cooperativity of the pump on the upper sideband. The colored points correspond to the curves on panel A. **C**. Relative gain and noise of the amplifier, sharing the same baseline. The difference from noise to gain corresponds to over 12 dB of apparent signal-to-noise ratio improvement of our device over the HEMT, from which the insertion loss between the HEMT and the device (measured separately to be 1.6 dB) must be subtracted to infer the real improvement. **D**. Added noise of the amplification expressed in quanta. The total added noise in the high-gain limit amounts to $1.68 \pm 0.02$ quanta, corroborated by an independent optomechanical calibration technique (cf. SI). This is only 0.87 quanta above the device quantum limit $n_\mathrm{DQL}$, defined in equation (10).

## ACKNOWLEDGMENTS

We thank A. A. Clerk, V. Sudhir and D. Wilson for helpful comments, E. Glushkov for helping out with the measurement setup and C. Javerzac-Galy for general assistance. This work was supported by the SNF, the NCCR Quantum Science and Technology (QSIT), the European Union Seventh Framework Program through iQUOEMS (grant no. 323924) and Marie Curie ITN cQOM (grant no. 290161). TJK acknowledges financial support from an ERC AdG (QuREM). AN holds a University Research Fellowship from the Royal Society and acknowledges support from the Winton Programme for the Physics of Sustainability. EG acknowledges financial support from ZKS-Foundation. All samples were fabricated in the Center of MicroNanoTechnology (CMi) at EPFL.

* L. D. Toth and N. R. Bernier contributed equally to this work.

† alexey.feofanov@epfl.ch

‡ tobias.kippenberg@epfl.ch

# Supplementary Information - A dissipative quantum reservoir for microwave light using a mechanical oscillator

L. D. Tóth,[1] N. R. Bernier,[1] A. Nunnenkamp,[2] A. K. Feofanov,[1,*] and T. J. Kippenberg[1,†]

[1]*Institute of Physics, École Polytechnique Fédérale de Lausanne (EPFL), CH-1015 Lausanne, Switzerland*
[2]*Cavendish Laboratory, University of Cambridge, Cambridge CB3 0HE, United Kingdom*

Supplementary Information accompanying the manuscript containing a theoretical analysis of the system dynamics as well as details on the experimental parameters, the fabrication of the device, the detailed measurement setup, and the calibration procedures.

## I. SYSTEM-RESERVOIR TREATMENT OF THE COUPLED CAVITY

In this section we give a theoretical description on how the dark and bright modes - used as primary and auxiliary modes - emerge from our circuit design. In particular, we present a system-reservoir treatment to describe the effect of intermode coupling on two degenerate electromagnetic modes coupled to a common bath (in our case, two electromagnetic modes inductively coupled to a common microwave feedline).

We start by considering two modes described by annihilation operators $\hat{a}_1$ and $\hat{a}_2$ that are coupled to each other with matrix element $J$ and that can dissipate into a common bath of harmonic oscillators described by their annihilation operators $\hat{b}_k$. The system-bath Hamiltonian describing this problem is

$$\begin{aligned}\hat{\mathcal{H}} &= \hbar\omega_0\left(\hat{a}_1^\dagger\hat{a}_1 + \hat{a}_2^\dagger\hat{a}_2\right) + \sum_k \hbar\omega_k \hat{b}_k^\dagger\hat{b}_k + \hbar J\left(\hat{a}_1^\dagger\hat{a}_2 + \text{H.c.}\right) \\ &+ \hbar\sum_k\left(g_k^{(1)}\hat{a}_1\hat{b}_k^\dagger + \text{H.c.}\right) + \hbar\sum_k\left(g_k^{(2)}\hat{a}_2\hat{b}_k^\dagger + \text{H.c.}\right). \qquad (1)\end{aligned}$$

Note that we take the coupling amplitudes $g_k^{(1)}$ and $g_k^{(2)}$ to be non-negative real numbers. In doing so, we will neglect a relative phase factor $e^{ikx}$ where $k$ is the wave vector of the microwaves and $x$ is the relative distance between the positions where the radio-frequency (RF) circuits are coupled to the common microwave feedline.

For $J \neq 0$ the degeneracy between the modes $\hat{a}_1$ and $\hat{a}_2$ is lifted and the normal modes are the symmetric and antisymmetric superpositions, $\hat{a}_{\text{s,a}} = \frac{1}{\sqrt{2}}\left(\hat{a}_1 \pm \hat{a}_2\right)$, with energies $\hbar(\omega_0 \pm J)$. In this basis, the Hamiltonian (1) reads

$$\begin{aligned}\hat{\mathcal{H}} &= \hbar(\omega_0 + J)\hat{a}_\text{s}^\dagger\hat{a}_\text{s} + \hbar(\omega_0 - J)\hat{a}_\text{a}^\dagger\hat{a}_\text{a} + \sum_k \hbar\omega_k\hat{b}_k^\dagger\hat{b}_k \qquad (2) \\ &+ \hbar\sum_k\left(\frac{g_k^{(1)}+g_k^{(2)}}{\sqrt{2}}\hat{a}_\text{s}\hat{b}_k^\dagger + \text{H.c.}\right) + \hbar\sum_k\left(\frac{g_k^{(1)}-g_k^{(2)}}{\sqrt{2}}\hat{a}_\text{a}\hat{b}_k^\dagger + \text{H.c.}\right).\end{aligned}$$

We see that if the coupling matrix elements are similar, i.e. $g_k^{(1)} \approx g_k^{(2)}$, the symmetric mode $\hat{a}_\text{s}$ features a much larger dissipation rate than that of the antisymmetric mode $\hat{a}_\text{a}$. Indeed, we have $g_k^{(+)} = (g_k^{(1)} + g_k^{(2)})/\sqrt{2} \approx \sqrt{2}g_k^{(1)}$ and $g_k^{(-)} = (g_k^{(1)} - g_k^{(2)})/\sqrt{2} \approx 0$ so that dissipation rate of the mode $\hat{a}_\text{s}$ has twice the dissipation rate of that of the mode $\hat{a}_1$, while the mode $\hat{a}_\text{a}$ is approximately decoupled from the common reservoir $\hat{b}_k$.

The situation is closely related to the case of a three-state $\Lambda$-type system in which the two transitions are driven coherently. In that case, so-called *bright* and *dark* states are formed. The population in the dark state is trapped and cannot decay even if the upper state of the $\Lambda$-type system has a finite linewidth.

In our case, the bright state $\hat{a}_\text{s}$ and the dark state $\hat{a}_\text{a}$ can be addressed separately by a coherent drive as long as their frequency difference $2J$ is large compared to their decay rates. In this situation, we can take $\hat{a}_\text{s}$ and $\hat{a}_\text{a}$ to be the modes of our system.

* alexey.feofanov@epfl.ch
† tobias.kippenberg@epfl.ch

## II. THEORETICAL BACKGROUND: GAIN, NOISE, AND OUTPUT SPECTRUM

In this section we give details about the model used to derive the expressions presented in the main text for amplifier gain, added noise, and device quantum limit as well as the output spectrum of our device.

Let us start by writing down the linearized equations of motion for our device consisting of one mechanical $\hat{b}$ and two optical modes, primary (anti-symmetric, dark) mode $\hat{a} = \hat{a}_{\mathrm{a}}$ and auxiliary (symmetric, bright) mode $\hat{a}_{\mathrm{aux}} = \hat{a}_{\mathrm{s}}$

$$\dot{\delta\hat{a}} = +i\Delta\delta\hat{a} - \frac{\kappa}{2}\delta\hat{a} + ig(\delta\hat{b} + \delta\hat{b}^\dagger) + \sqrt{\kappa_{\mathrm{ex}}}\,\delta\hat{a}_{\mathrm{in,ex}} + \sqrt{\kappa_0}\,\delta\hat{a}_{\mathrm{in,0}} \tag{3}$$

$$\dot{\delta\hat{b}} = -i\Omega_{\mathrm{m}}\delta\hat{b} - \frac{\Gamma_{\mathrm{m}}}{2}\delta\hat{b} + ig(\delta\hat{a} + \delta\hat{a}^\dagger) + ig_{\mathrm{aux}}(\delta\hat{a}_{\mathrm{aux}} + \delta\hat{a}^\dagger_{\mathrm{aux}}) + \sqrt{\Gamma_{\mathrm{m}}}\,\delta\hat{b}_{\mathrm{in}} \tag{4}$$

$$\dot{\delta\hat{a}}_{\mathrm{aux}} = +i\Delta_{\mathrm{aux}}\delta\hat{a}_{\mathrm{aux}} - \frac{\kappa_{\mathrm{aux}}}{2}\delta\hat{a}_{\mathrm{aux}} + ig_{\mathrm{aux}}(\delta\hat{b} + \delta\hat{b}^\dagger) + \sqrt{\kappa^{\mathrm{ex}}_{\mathrm{aux}}}\,\delta\hat{a}^{(\mathrm{aux})}_{\mathrm{in,ex}} + \sqrt{\kappa^{(0)}_{\mathrm{aux}}}\,\delta\hat{a}^{(\mathrm{aux})}_{\mathrm{in,0}} \tag{5}$$

with the detunings $\Delta = \omega_L - \omega_c$ and $\Delta_{\mathrm{aux}} = \omega_L^{\mathrm{aux}} - \omega_{\mathrm{aux}}$, the drive frequencies $\omega_L$ and $\omega_L^{\mathrm{aux}}$, the electromechanical coupling strengths $g = g_0|\bar{a}|$ and $g_{\mathrm{aux}} = g_0^{\mathrm{aux}}|\bar{a}_{\mathrm{aux}}|$ enhanced by the mean number of photons $|\bar{a}|^2$ and $|\bar{a}_{\mathrm{aux}}|^2$, and the input noise operators which satisfy bosonic commutation relations as well as

$$\langle\delta\hat{a}_{\mathrm{in,ex}}(t)\delta\hat{a}^\dagger_{\mathrm{in,ex}}(t')\rangle = (n_{\mathrm{in}} + 1)\delta(t - t') \tag{6}$$

$$\langle\delta\hat{a}_{\mathrm{in,0}}(t)\delta\hat{a}^\dagger_{\mathrm{in,0}}(t')\rangle = (n_{\mathrm{cav}} + 1)\delta(t - t') \tag{7}$$

$$\langle\delta\hat{b}_{\mathrm{in}}(t)\delta\hat{b}^\dagger_{\mathrm{in}}(t')\rangle = (n_{\mathrm{th}} + 1)\delta(t - t') \tag{8}$$

$$\langle\delta\hat{a}^{(\mathrm{aux})}_{\mathrm{in,ex}}(t)\delta\hat{a}^{(\mathrm{aux})\dagger}_{\mathrm{in,ex}}(t')\rangle = (n^{(\mathrm{aux})}_{\mathrm{in}} + 1)\delta(t - t') \tag{9}$$

$$\langle\delta\hat{a}^{(\mathrm{aux})}_{\mathrm{in,0}}(t)\delta\hat{a}^{(\mathrm{aux})\dagger}_{\mathrm{in,0}}(t')\rangle = (n^{(\mathrm{aux})}_{\mathrm{cav}} + 1)\delta(t - t') \tag{10}$$

which describe thermal fluctuations of the external feedlines and cavity modes as well as the mechanical oscillator.

In the rotating-wave approximation $\kappa, \kappa_{\mathrm{aux}} \ll \Omega_{\mathrm{m}}$ this set of equations simplifies considerably. Pumping the primary mode close to the blue sideband $\Delta \approx \Omega_{\mathrm{m}}$ and the auxiliary mode close to the red sideband $\Delta_{\mathrm{aux}} \approx -\Omega_{\mathrm{m}}$, we obtain

$$\dot{\delta\hat{a}} = +i\Delta\delta\hat{a} - \frac{\kappa}{2}\delta\hat{a} + ig\delta\hat{b}^\dagger + \sqrt{\kappa_{\mathrm{ex}}}\,\delta\hat{a}_{\mathrm{in,ex}} + \sqrt{\kappa_0}\,\delta\hat{a}_{\mathrm{in,0}} \tag{11}$$

$$\dot{\delta\hat{b}}^\dagger = +i\Omega_{\mathrm{m}}\delta\hat{b}^\dagger - \frac{\Gamma_{\mathrm{m}}}{2}\delta\hat{b}^\dagger - ig\delta\hat{a} - ig_{\mathrm{aux}}\delta\hat{a}^\dagger_{\mathrm{aux}} + \sqrt{\Gamma_{\mathrm{m}}}\,\delta\hat{b}^\dagger_{\mathrm{in}} \tag{12}$$

$$\dot{\delta\hat{a}}^\dagger_{\mathrm{aux}} = -i\Delta_{\mathrm{aux}}\delta\hat{a}^\dagger_{\mathrm{aux}} - \frac{\kappa_{\mathrm{aux}}}{2}\delta\hat{a}^\dagger_{\mathrm{aux}} - ig_{\mathrm{aux}}\delta\hat{b}^\dagger + \sqrt{\kappa^{\mathrm{ex}}_{\mathrm{aux}}}\,\delta\hat{a}^{(\mathrm{aux})\dagger}_{\mathrm{in,ex}} + \sqrt{\kappa^{(0)}_{\mathrm{aux}}}\,\delta\hat{a}^{(\mathrm{aux})\dagger}_{\mathrm{in,0}}. \tag{13}$$

The auxiliary mode can be eliminated from this set of equations

$$\dot{\delta\hat{a}} = +i\Delta\delta\hat{a} - \frac{\kappa}{2}\delta\hat{a} + ig\delta\hat{b}^\dagger + \sqrt{\kappa_{\mathrm{ex}}}\,\delta\hat{a}_{\mathrm{in,ex}} + \sqrt{\kappa_0}\,\delta\hat{a}_{\mathrm{in,0}} \tag{14}$$

$$\dot{\delta\hat{b}}^\dagger = +i\Omega_{\mathrm{m}}\delta\hat{b}^\dagger - \frac{\Gamma_{\mathrm{eff}}}{2}\delta\hat{b}^\dagger - ig\delta\hat{a} + \sqrt{\Gamma_{\mathrm{eff}}}\,\delta\hat{b}^\dagger_{\mathrm{in}}. \tag{15}$$

where for $\Delta_{\mathrm{aux}} = -\Omega_{\mathrm{m}}$ we introduced an effective mechanical linewidth $\Gamma_{\mathrm{eff}} = \Gamma_{\mathrm{m}} + \frac{4g^2_{\mathrm{aux}}}{\kappa_{\mathrm{aux}}}$ and input noise correlators

$$\langle\delta\hat{b}_{\mathrm{in}}(t)\delta\hat{b}^\dagger_{\mathrm{in}}(t')\rangle = (n_{\mathrm{eff}} + 1)\delta(t - t') \tag{16}$$

with an effective mechanical occupation

$$n_{\mathrm{eff}} = \frac{1}{\Gamma_{\mathrm{eff}}}\left[\Gamma_{\mathrm{m}}n_{\mathrm{th}} + \frac{4g^2_{\mathrm{aux}}}{\kappa^2_{\mathrm{aux}}}\left(\kappa^{\mathrm{ex}}_{\mathrm{aux}}n^{(\mathrm{aux})}_{\mathrm{in}} + \kappa^{(0)}_{\mathrm{aux}}n^{(\mathrm{aux})}_{\mathrm{cav}}\right)\right]. \tag{17}$$

In this approximation equations (14) and (15) are those of an non-degenerate parametric amplifier which can be solved in terms of the mechanical response function $\chi_{\mathrm{eff}}[\omega] = [\Gamma_{\mathrm{eff}}/2 - i(\omega - \Omega_{\mathrm{m}})]^{-1}$

$$\delta\hat{b}^\dagger[\omega] = \chi^\star_{\mathrm{eff}}[-\omega]\left[\sqrt{\Gamma_{\mathrm{eff}}}\,\delta\hat{b}^\dagger_{\mathrm{in}}[\omega] - ig\delta\hat{a}[\omega]\right] \tag{18}$$

and

$$\left[\frac{\kappa}{2} - i(\omega + \Delta) - g^2\chi^\star_{\mathrm{eff}}[-\omega]\right]\delta\hat{a}[\omega] = \sqrt{\kappa_{\mathrm{ex}}}\,\delta\hat{a}_{\mathrm{in,ex}}[\omega] + \sqrt{\kappa_0}\,\delta\hat{a}_{\mathrm{in,0}}[\omega] + ig\chi^\star_{\mathrm{eff}}[-\omega]\sqrt{\Gamma_{\mathrm{eff}}}\,\delta\hat{b}^\dagger_{\mathrm{in}}[\omega]. \tag{19}$$

For $\Delta = \Omega_\mathrm{m}$, we have $\delta\omega_\mathrm{om} = 0$ and $\kappa_\mathrm{om} = -\frac{4g^2}{\Gamma_\mathrm{eff}}$. For $\kappa \ll \Gamma_\mathrm{eff}$ we can neglect the frequency dependence of $\chi_\mathrm{eff}[\omega]$

$$\left[\frac{\kappa_\mathrm{ex} + \kappa_0 + \kappa_\mathrm{om}}{2} - i(\omega + \Delta)\right] \delta\hat{a}[\omega] = \sqrt{\kappa_\mathrm{ex}}\,\delta\hat{a}_\mathrm{in,ex}[\omega] + \sqrt{\kappa_0}\,\delta\hat{a}_\mathrm{in,0}[\omega] + i\sqrt{\mathcal{C}\kappa}\,\delta\hat{b}^\dagger_\mathrm{in}[\omega] \tag{20}$$

making explicit that the cavity mode $\hat{a}$ is coupled to three different reservoirs: to the external feedline at rate $\kappa_\mathrm{ex}$, to internal dissipation at rate $\kappa_0$, and to the mechanical oscillator at rate $\mathcal{C}\kappa$.

Using the input-output relation $\delta\hat{a}_\mathrm{out} = \delta\hat{a}_\mathrm{in,ex} - \sqrt{\kappa_\mathrm{ex}}\,\delta\hat{a}$ and defining a modified cavity response function as

$$\tilde{\chi}_R[\omega] = \left[\frac{\kappa}{2} - i(\omega + \Delta) - g^2\chi^\star_\mathrm{eff}[-\omega]\right]^{-1}, \tag{21}$$

we express the fluctuations at the output in terms of the fluctuations at the input

$$\delta\hat{a}_\mathrm{out}[\omega] = A(\omega)\delta\hat{a}_\mathrm{in,ex}[\omega] + B(\omega)\delta\hat{a}_\mathrm{in,0}[\omega] + C(\omega)\delta\hat{b}^\dagger_\mathrm{in}[\omega] \tag{22}$$

with

$$A(\omega) = 1 - \kappa_\mathrm{ex}\tilde{\chi}_R[+\omega] \tag{23}$$

$$B(\omega) = -\sqrt{\kappa_\mathrm{ex}\kappa_0}\tilde{\chi}_R[+\omega] \tag{24}$$

$$C(\omega) = -ig\sqrt{\kappa_\mathrm{ex}\Gamma_\mathrm{eff}}\tilde{\chi}_R[+\omega]\chi^\star_\mathrm{eff}[-\omega] \tag{25}$$

satisfying $|A(\omega)|^2 + |B(\omega)|^2 - |C(\omega)|^2 = 1$. For consistency, we note that our notation of $S_{11}(\omega)$ in the main text corresponds to $A(\omega)$ in this derivation.

The symmetrized cavity output field spectrum is given by

$$\bar{S}_{aa}(\omega) = \frac{1}{2}\int_{-\infty}^{\infty} dt\, e^{i\omega t}\langle \delta\hat{a}^\dagger_\mathrm{out}(t)\delta\hat{a}_\mathrm{out}(0) + \delta\hat{a}_\mathrm{out}(0)\delta\hat{a}^\dagger_\mathrm{out}(t)\rangle \tag{26}$$

$$= |A(-\omega)|^2(n_\mathrm{in} + \tfrac{1}{2}) + |B(-\omega)|^2(n_\mathrm{cav} + \tfrac{1}{2}) + |C(-\omega)|^2(n_\mathrm{eff} + \tfrac{1}{2}) \tag{27}$$

$$= \left|1 - \kappa_\mathrm{ex}\tilde{\chi}_R[-\omega]\right|^2(n_\mathrm{in} + \tfrac{1}{2}) + \kappa_\mathrm{ex}\kappa_0|\tilde{\chi}_R[-\omega]|^2(n_\mathrm{cav} + \tfrac{1}{2}) + |g|^2\kappa_\mathrm{ex}\Gamma_\mathrm{eff}|\tilde{\chi}_R[-\omega]|^2|\chi_\mathrm{eff}[+\omega]|^2(n_\mathrm{eff} + \tfrac{1}{2}). \tag{28}$$

For $\Gamma_\mathrm{eff} \gg \kappa_\mathrm{eff} = \kappa_\mathrm{ex} + \kappa_0 + \kappa_\mathrm{om}$ and $\Delta = \Omega_\mathrm{m}$ we obtain in the lab frame

$$\bar{S}_{aa}(\omega) = \left[1 + \frac{\kappa_\mathrm{ex}(\kappa_\mathrm{ex} - \kappa_\mathrm{eff})}{(\frac{\kappa_\mathrm{eff}}{2})^2 + (\omega - \omega_c)^2}\right](n_\mathrm{in} + \tfrac{1}{2}) + \frac{\kappa_\mathrm{ex}\kappa_0}{(\frac{\kappa_\mathrm{eff}}{2})^2 + (\omega - \omega_c)^2}(n_\mathrm{cav} + \tfrac{1}{2}) + \frac{\mathcal{C}\kappa_\mathrm{ex}\kappa}{(\frac{\kappa_\mathrm{eff}}{2})^2 + (\omega - \omega_c)^2}(n_\mathrm{eff} + \tfrac{1}{2}). \tag{29}$$

We see that the output spectrum $\bar{S}_{aa}(\omega)$ can be understood as amplified noise (thermal and zero-point fluctuations) from the three reservoirs. Indeed, with the expressions for gain and added noise below we obtain in the rotating frame

$$\bar{S}_{aa}(\omega) = \mathcal{G}(-\omega)\left[n_\mathrm{in} + \tfrac{1}{2} + \mathcal{N}(-\omega)\right]. \tag{30}$$

The power gain of the amplifier is $\mathcal{G}(\omega) = |A(\omega)|^2$ and for pump on sideband $\Delta = \Omega_\mathrm{m}$ signal on resonance $\omega = -\Delta_\mathrm{eff}$

$$\mathcal{G}(\omega_c) = \left(1 - \frac{2\kappa_\mathrm{ex}}{\kappa_\mathrm{eff}}\right)^2 = \left(\frac{1 - \mathcal{C} - 2\frac{\kappa_\mathrm{ex}}{\kappa}}{1 - \mathcal{C}}\right)^2. \tag{31}$$

and the added noise of the amplifier is

$$\mathcal{N}(\omega) = (n_\mathrm{cav} + \tfrac{1}{2})\left|\frac{B(\omega)}{A(\omega)}\right|^2 + (n_\mathrm{eff} + \tfrac{1}{2})\left|\frac{C(\omega)}{A(\omega)}\right|^2 \tag{32}$$

and for pump on sideband $\Delta = \Omega_\mathrm{m}$ and signal on resonance $\omega = -\Delta_\mathrm{eff}$ we obtain

$$\mathcal{N}(\omega_c) = \frac{\kappa_\mathrm{ex}\kappa_0(n_\mathrm{cav} + \frac{1}{2})}{(\frac{\kappa_\mathrm{eff}}{2} - \kappa_\mathrm{ex})^2} + \frac{\mathcal{C}\kappa_\mathrm{ex}\kappa(n_\mathrm{eff} + \frac{1}{2})}{(\frac{\kappa_\mathrm{eff}}{2} - \kappa_\mathrm{ex})^2} = \frac{4\frac{\kappa_\mathrm{ex}\kappa_0}{\kappa^2}(n_\mathrm{cav} + \frac{1}{2})}{(1 - \mathcal{C} - 2\frac{\kappa_\mathrm{ex}}{\kappa})^2} + \frac{4\mathcal{C}\frac{\kappa_\mathrm{ex}}{\kappa}(n_\mathrm{eff} + \frac{1}{2})}{(1 - \mathcal{C} - 2\frac{\kappa_\mathrm{ex}}{\kappa})^2}. \tag{33}$$

At small gain $\mathcal{C} \to 0$ the added noise is mostly intracavity noise, in particular close to critical coupling, i.e. $\kappa_\mathrm{ex} \approx \kappa_0$, where the gain goes to zero, $\mathcal{G}(\omega_c) \to 0$, since the signal is perfectly absorbed by the internal dissipation of the cavity,

$$\mathcal{N}(\omega_c) \to \frac{4\kappa_\mathrm{ex}\kappa_0(n_\mathrm{cav} + \frac{1}{2})}{(\kappa_\mathrm{ex} - \kappa_0)^2} = \frac{4\frac{\kappa_0}{\kappa_\mathrm{ex}}(n_\mathrm{cav} + \frac{1}{2})}{(1 - \frac{\kappa_0}{\kappa_\mathrm{ex}})^2}. \tag{34}$$

Due to finite internal dissipation the added noise of the amplifier for large gain $\mathcal{C} \to 1$ becomes

$$\mathcal{N}(\omega_c) \to \frac{\kappa_0}{\kappa_{\rm ex}}(n_{\rm cav} + \tfrac{1}{2}) + \frac{\kappa}{\kappa_{\rm ex}}(n_{\rm eff} + \tfrac{1}{2}) = \tfrac{1}{2} + n_{\rm eff} + \frac{\kappa_0}{\kappa_{\rm ex}}(n_{\rm eff} + n_{\rm cav} + 1) > \tfrac{1}{2}. \tag{35}$$

In the main manuscript we refer to $n_{\rm DQL} = \frac{1}{2} + \frac{\kappa_0}{\kappa_{\rm ex}}$ as the *device quantum limit* for the added noise of our amplifier due to the presence of a finite coupling to the output feedline with rate $\kappa_{\rm ex}$ compared to an intrinsic damping rate $\kappa_0$.

## III. DEVICE PARAMETERS AND CIRCUIT DESIGN

Table I. and II. summarizes the most important parameters of the devices measured (the amplifier results were measured using device 2). As mentioned in the main text and in Sec. I of the Supplementary Information, in

TABLE I. Summary of the most important experimental parameters, device 1.

| **Parameter** | **Symbol** | **Value** |
|---|---|---|
| Resonance frequency (primary, dark mode) | $\omega_{\rm c}$ | $2\pi \cdot 4.26$ GHz |
| Total decay rate (primary, dark mode) | $\kappa$ | $2\pi \cdot 118$ kHz |
| Internal decay rate (primary, dark mode) | $\kappa_0$ | $2\pi \cdot 76$ kHz |
| External coupling rate (primary, dark mode) | $\kappa_{\rm ex}$ | $2\pi \cdot 42$ kHz |
| Resonance frequency (auxiliary, bright mode) | $\omega_{\rm aux}$ | $2\pi \cdot 5.48$ GHz |
| Total decay rate (auxiliary, bright mode) | $\kappa_{\rm aux}$ | $2\pi \cdot 4478$ kHz |
| Internal decay rate (auxiliary, bright mode) | $\kappa_{\rm aux}^{(0)}$ | $2\pi \cdot 245$ kHz |
| External coupling rate (auxiliary, bright mode) | $\kappa_{\rm aux}^{\rm ex}$ | $2\pi \cdot 4233$ kHz |
| Intermode coupling strength (also estimated from simulations) | $J$ | $2\pi \cdot 0.57$ GHz |
| Resonance frequency of mechanical mode | $\Omega_{\rm m}$ | $2\pi \cdot 5.33$ MHz |
| Decay rate of mechanical mode | $\Gamma_{\rm m}$ | $2\pi \cdot 30$ Hz |
| Effective decay rate of mechanical mode in the RDR | $\Gamma_{\rm eff}$ | $2\pi \cdot 500$ kHz |
| Vacuum electromechanical coupling strength (for the hybridized modes) | $g_0 = \tilde{g}_0/2$ | $2\pi \cdot 60$ Hz |

TABLE II. Summary of the most important experimental parameters, device 2.

| **Parameter** | **Symbol** | **Value** |
|---|---|---|
| Resonance frequency (primary, dark mode) | $\omega_{\rm c}$ | $2\pi \cdot 4.13$ GHz |
| Total decay rate (primary, dark mode) | $\kappa$ | $2\pi \cdot 197$ kHz |
| Internal decay rate (primary, dark mode) | $\kappa_0$ | $2\pi \cdot 47$ kHz |
| External coupling rate (primary, dark mode) | $\kappa_{\rm ex}$ | $2\pi \cdot 150$ kHz |
| Resonance frequency (auxiliary, bright mode) | $\omega_{\rm aux}$ | $2\pi \cdot 5.24$ GHz |
| Total decay rate (auxiliary, bright mode) | $\kappa_{\rm aux}$ | $2\pi \cdot 3322$ kHz |
| Internal decay rate (auxiliary, bright mode) | $\kappa_{\rm aux}^{(0)}$ | $2\pi \cdot 250$ kHz |
| External coupling rate (auxiliary, bright mode) | $\kappa_{\rm aux}^{\rm ex}$ | $2\pi \cdot 3072$ kHz |
| Intermode coupling strength ($\approx \lvert\omega_{\rm aux} - \omega_{\rm c}\rvert/2$) | $J$ | $2\pi \cdot 0.56$ GHz |
| Resonance frequency of mechanical mode | $\Omega_{\rm m}$ | $2\pi \cdot 6.35$ MHz |
| Decay rate of the mechanical mode | $\Gamma_{\rm m}$ | $2\pi \cdot 100$ Hz |
| Effective decay rate of mechanical mode in the RDR | $\Gamma_{\rm eff}$ | $2\pi \cdot 400$ kHz |
| Vacuum electromechanical coupling strength (primary and auxiliary mode) | $g_0, g_0^{\rm aux}$ | $2\pi \cdot (106, 79)$ Hz |

order to realize a hybrid-mode circuit featuring a bright and a dark mode, we need to engineer the individual circuit parameters so that their bare resonance frequencies match. Additionally, we need to strongly couple them (i.e. $J$ is engineered to be much greater than the energy decay rates of the bare modes and the difference of their frequencies), which ensures that they can be individually addressed and that the normal-mode splitting occurs even if there are deviations in their bare resonance frequencies. Our layout consists of two LC resonators sharing an inductor line that are both inductively coupled to a common feedline. One LC resonator contains the mechanically compliant drum-type capacitor, the other one contains an interdigitated capacitor. We simulate both the total inductances (including kinetic inductance) and capacitances for the circuits separately to aid the design, using commercially available tools (Sonnet and FastHenry). Fig. S1 shows the simulated current distribution of the modes. Table III contains some of the results from simulation for the measured device geometry.

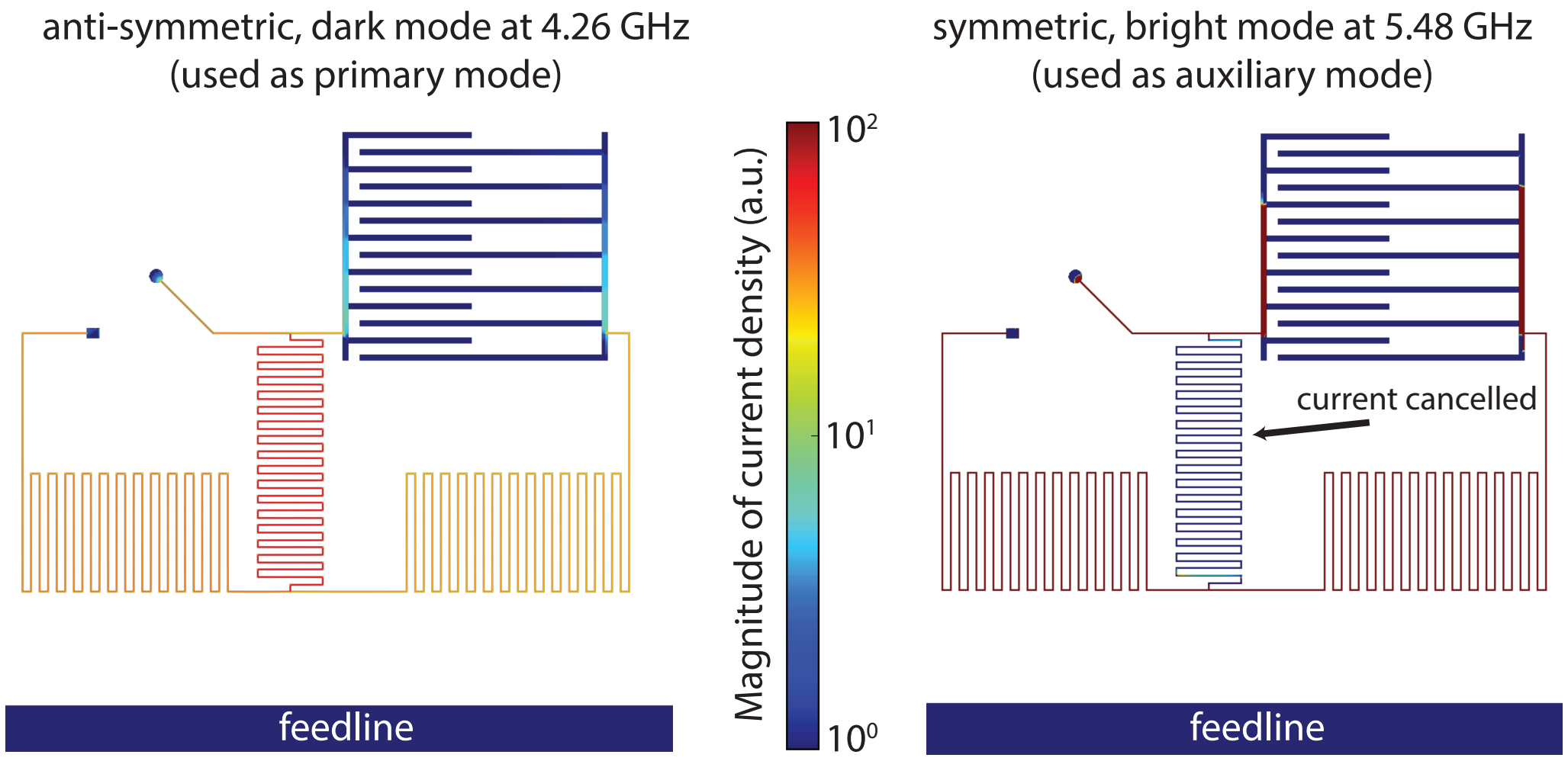

Fig. S 1. Simulation of the circuit revealing the symmetric and anti-symmetric modes. The symmetric mode has no current on the coupling inductor (meandering line in the middle). The anti-symmetric mode has current flowing in the opposite direction on the lower inductors, therefore its external coupling to the feedline is suppressed.

TABLE III. Simulated parameters and comparison to experimental results

| Parameter | Simulated value |
|---|---|
| $\omega_c$ | $2\pi \cdot 4.348$ GHz ($< 2\%$ deviation from measurements) |
| $\omega_{\mathrm{aux}}$ | $2\pi \cdot 5.482$ GHz ($< 2\%$ deviation from measurements) |
| Inductance of individual LC circuit | $\approx 11$ nH |
| Capacitance of the interdigitated capacitor | $\approx 104$ fF |
| Inferred plate separation of vacuum-gap capacitor | $\approx 38$ nm |

## IV. FABRICATION OF THE DEVICE

We fabricate the device on a standard 525 $\mu$m thick C-plane sapphire wafer using standard photolithography techniques (the key ideas of the fabrication process are similar to the ones developed in Cicak *et al.* [1], with differences in the etch chemistries and sacrificial layer material). Fig. S2 shows the entire process flow. After initial surface preparation - consisting of a full RCA cleaning and annealing at 1000 °C for 1 hour - we evaporate 100 nm of aluminium at room temperature. We pattern this layer by direct laser lithography and Cl-based dry etching. Then, 150 nm of amorphous Si layer is DC Magnetron sputtered which will serve as a sacrificial layer for the suspended electrode and also protects the rest of the circuit during the process. The sacrificial layer is patterned in two steps, one with a F-based chemistry and one with a Cl-based chemistry (the former is used to open the Si layer above the connecting Al pad, the latter is used where the top electrode connects to the substrate and, by using a reflown resist, achieves slanted sidewalls). Next, we evaporate another layer of aluminium (100 nm) at room temperature, which will form the suspended top electrode. This layer is patterned similarly as the first Al layer. The top Al layer connects to the bottom layer by a large connecting pad. Although this pad has native oxide on its surface before the top layer Al evaporation (preventing galvanic contact), it forms a capacitor with capacitance much larger than the vacuum-gap capacitance, rendering it irrelevant. After the top Al layer is patterned, we dice the wafer into chips followed by careful cleaning. The last step is to release the top electrode, which we achieve using a $XeF_2$ dry process.

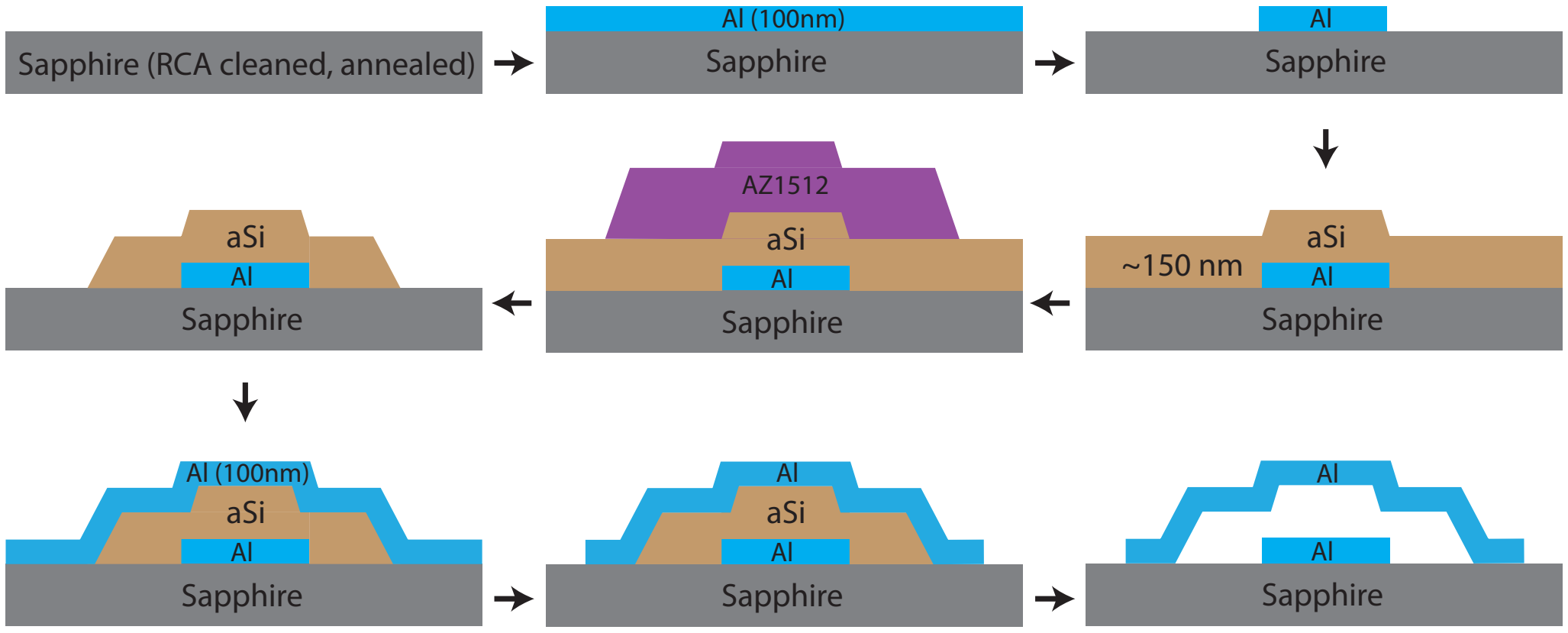

Fig. S 2. Process flow for fabricating the superconducting electromechanical circuit. The process employs 3 deposition-lithography-etching steps to define the circuit, followed by the release of the drum-type capacitor.

## V. DETAILED DESCRIPTION OF THE MEASUREMENT SETUP

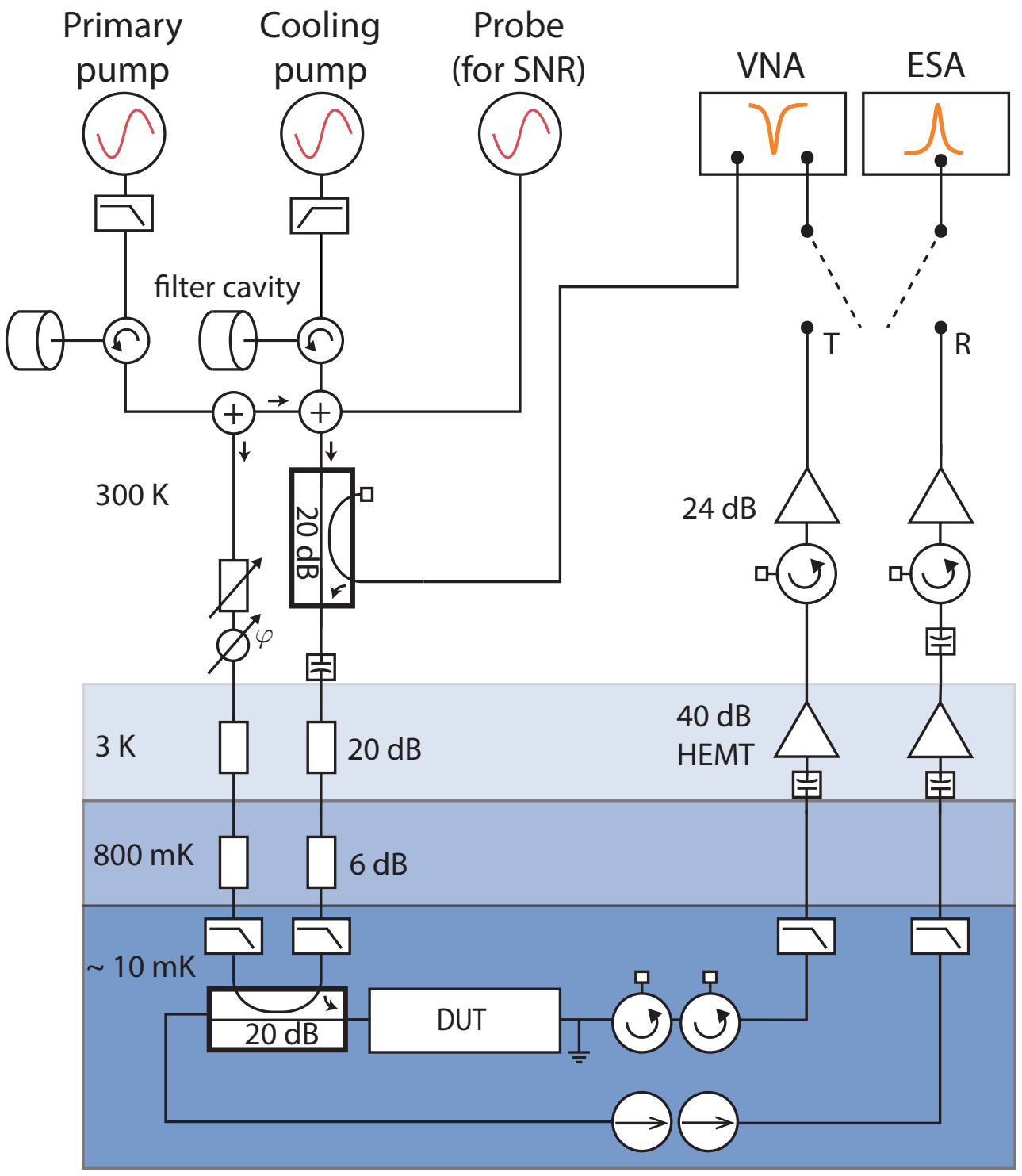

Fig. S 3. Schematics of the experimental setup. See text for description and details on the components used.

The detailed diagram of the measurement setup is shown on Fig. S3. We combine microwave sources (R&S SMB and SMF) at room temperature together with the output of a vector network analyser. The cooling pump is used as the cooling/damping tone on the auxiliary microwave mode, while the primary pump is used on the primary microwave mode (cf. main text). Both of these tones are filtered at multiple stages. First, the primary pump and the cooling pump are connected to low-pass and high-pass filters, respectively (importantly it was observed that the sources emit an observable amount of noise *far away* from the carrier frequency). Additionally, they are both filtered using custom-made, tunable copper cylindrical cavities (similar in design to the one used by Lecocq *et al.* [2]) to improve their phase-noise performance around the relevant frequencies (see Fig. S4). The (weak) probe tone is used only for the measurements when our device is operated as an amplifier, to characterize the signal-to-noise ratio (cf. fig 4 E in the main text and Fig. S8). Before the primary and cooling pumps are combined with the probe tone, they are split and half of each one is sent through a variable attenuator and phase shifter (on different lines for the two pumps; for simplicity, on the primary pump line is shown on the figure). These lines are later combined with the reflection output line to suppress by interference the strong primary and cooling pumps reflected from the sample (thus preventing compression of the HEMT amplifier on the reflection channel). We could reach suppression factors of over 30 dB.

The input line with the combined tones reaches the sample through various stages of attenuation to suppress Johnson-noise (total attenuation in the dilution refrigerator, including cables and connections, of about 50 dB). Low-pass filters (K&L tubular low pass 0-12 GHz) are also employed on all input and output lines to filter out any unwanted high-frequency radiation. The sample itself is glued and wire-bonded in a sample holder made of oxygen-free high-conductivity copper, which is mounted on the base plate of the dilution refrigerator. One port of the sample holder is grounded; however, it is still connected to an output line ("T" line on the schematics) and can be measured for the characterization of the cavities resonances (at a level 20 dB below the reflected signal; cf. section VI). The transmission signal first goes through two stages of isolation at base temperature (using Raditek 4-8 GHz circulators), then a HEMT amplifier (LNF-LNC4-16A 4-16 GHz cryogenic low noise amplifier from Low Noise Factory) at 3 K. Finally, at room temperature, it is isolated and amplified again (isolator: Narda 2-stage isolator; amplifier: Minicircuit ZVA183+ 24-26 dB). Most of the signal from the device is reflected and directed to the reflection line ("R" line on the schematics) with a directional coupler (Pasternak PE2204-20 4-8 GHz). This line goes through two stages of isolation at base temperature (using Quinstar 4-12 GHz isolators), then a HEMT amplifier (LNF-LNC1-12 A 1-12 GHz cryogenic low noise amplifier from Low Noise Factory) at 3 K. Finally, at room temperature, it is isolated and amplified again (isolator: Innowave 2-stage isolator; amplifier: 2 in-series Minicircuit ZX60-14012L-S+ 12 dB). All the lines between the 3 K stage and the mixing chamber are wired using NbTi 0.86 mm superconducting cables, while formable Cu cables are used at the mixing chamber stage. Both the reflection and transmission lines can be measured with a vector network analyser (R&S ZNB20) or the electromagnetic spectrum analyser (R&S FSW26).

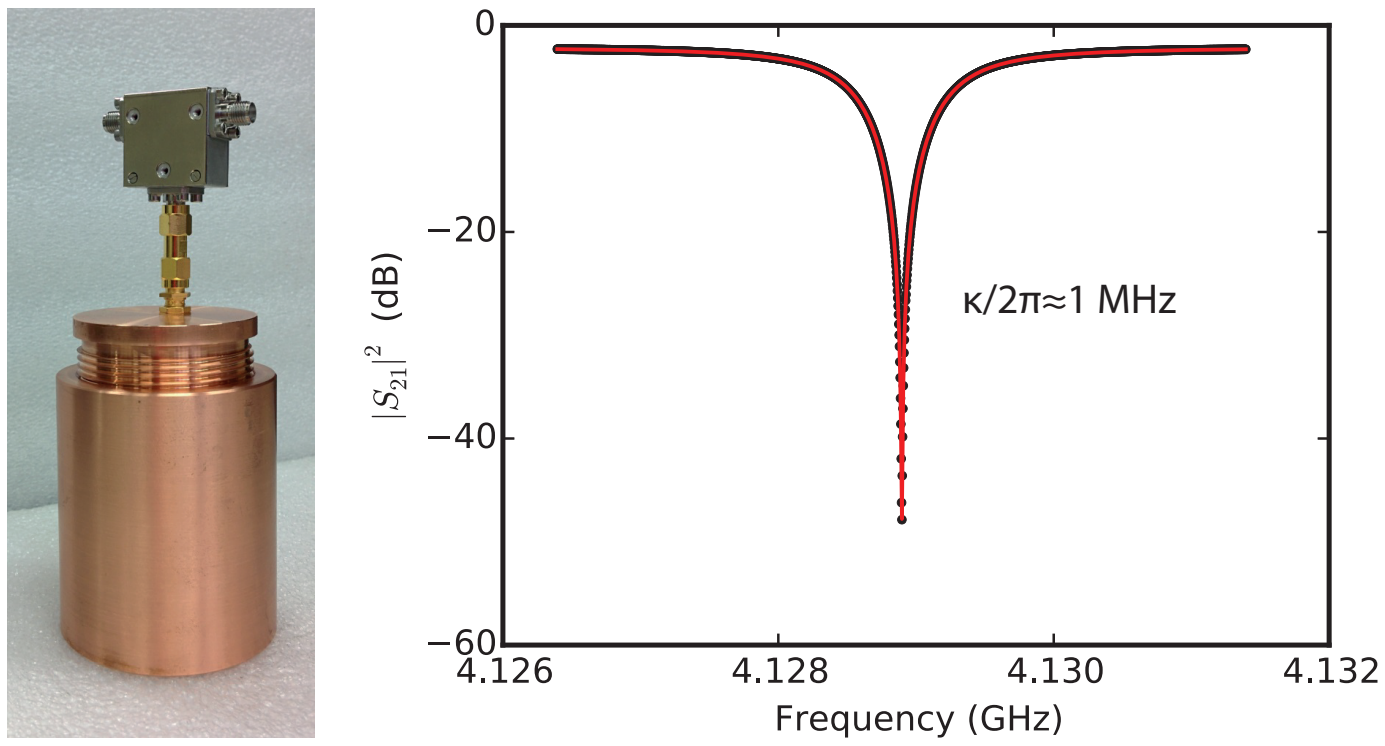

Fig. S 4. Left panel: photograph of the tunable copper cylindrical cavity used for reducing the phase noise on the microwave tones. Right panel: linear response of the cavity connected to a vector network analyzer in a notch-type configuration.

## VI. EXTRACTING PARAMETERS OF THE MICROWAVE MODE

In order to characterize the microwave cavities and extract the resonant frequencies, external coupling and internal energy decay rate, we use a method of fitting the complex linear response of the microwave resonance with a circle. We can use both the transmission and the reflection responses in order to characterize the resonance.

Following the analysis by Megrant *et al.* [3], the transmission response of a cavity of resonance frequency $\omega_0$ in a notch configuration with impedance mismatch of the two output channels is given by

$$S_{21}(\omega) = A\left(1 + \frac{Q_{\rm i}}{Q_{\rm c}^*}e^{i\varphi}\frac{1}{1+i2Q_{\rm i}\delta x}\right)^{-1} \tag{36}$$

with $\delta x = \frac{\omega-\omega_0}{\omega_0}$, $Q_i = \frac{\omega_0}{\kappa_0}$ the internal quality factor, and $\varphi$ a complex phase depending on the impedance mismatch of the channels. The external quality factor $Q_{\rm c} = \frac{\omega_0}{\kappa_{\rm ex}}$ is given by $Q_{\rm c} = Q_{\rm c}^*/\cos(\varphi)$ [4]. The constant $A$ includes all absorption and complex phase on the output line.

The inverse of $S_{21}(\omega)$ describes a circle in the complex plane. It can be fitted to extract all the parameters. We use the method of Probst *et al.* [5] and adapt their code to fit circles to our data. First, the constant $A$ can be normalized away by using the fact that $S_{21}$ reduces to $A$ far from resonance (i.e. for large $\delta x$). Then, the angle of rotation around the circle as a function of frequency can be fitted to extract $\omega_0$ and $Q_{\rm i}$. Finally, the impedance mismatch phase $\varphi$ can be extracted from the position of the center of the circle and the ratio $Q_{\rm i}/Q_{\rm c}^*$ from the circle diameter.

Reflection measurements can also be fitted using the circle method. In that case, the term of 1 in eq. (36) is lacking, with the consequence that there is no possibility to extract both $Q_{\rm i}$ and $Q_{\rm c}$. A circle fit can still be made for $S_{11}$, and the full width $\kappa_{\rm eff}$ and the resonance frequency can be extracted from the angle of rotation around the circle (see Fig. S5 for an example).

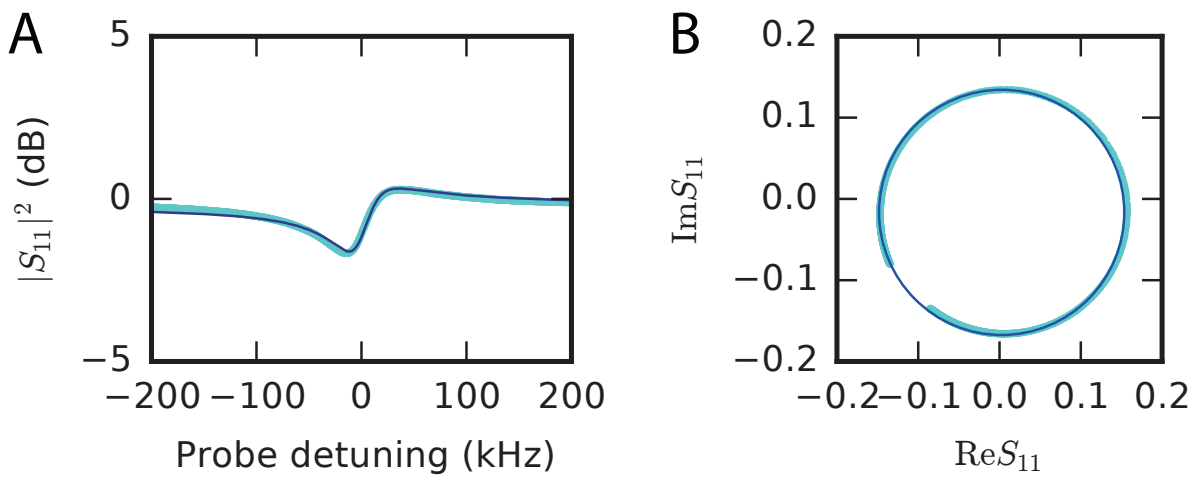

Fig. S 5. An example of fitting the linear response in reflection ($S_{11}$) with a circle on the complex plane, for a pump tone on the upper motional sideband such that the response is almost flat in amplitude and shows a strong Fano shape. The data corresponds to the turquoise curve of Figure 3 B in the main text.

## VII. CALIBRATIONS

We describe here the different calibration procedures used to extract the parameters of the system, as performed for device 2.

### A. Thermalization of the mechanical mode

In order to measure the vacuum coupling rate $g_0$, one must have an absolute reference of energy, here chosen to be the intrinsic thermal fluctuations of the mechanical oscillator. One must first show that the oscillator is in thermal equilibrium with the base plate of the refrigerator, and that the thermometer temperature can be used as an absolute measurement of the oscillator temperature. We measure the blue sideband imparted by the mechanics to a weak probe on resonance with the cavity (see Fig. S6 B). The integrated power in the Lorentzian peak (in the high temperature limit $n_{\rm m} \gg 1$) is given by [6]

$$P_{\rm SB} = \hbar\omega\frac{\kappa_{\rm ex}}{\Omega_{\rm m}^2 + (\kappa/2)^2}g_0^2 n_{\rm c} n_{\rm m} \tag{37}$$

where $\omega$ is the frequency of the mode, $n_{\rm c}$ is the mean cavity occupation due to the probe, and $n_{\rm m}$ is the mean mechanical phonon occupation. For fixed power and detuning of the probe, the sideband power is directly proportional to $n_{\rm m}$. The sideband power was measured at different temperature of the base plate (adjusted using a heater), with the result shown in Fig. S6 A. For the oscillator undamped, we did observe a perturbation from the pulse tube of the dilution refrigerator, acting as an additional noise source (causing frequency jitter for the mechanical and microwave resonances). As a precaution, the pulse tube was switched off for the duration of the measurement of the sideband. At high temperature, $n_{\rm m}(T)$ follows linearly the temperature of the base plate, so that we can assume it is thermalized. The power data is then normalized to $n_{\rm m}$ by anchoring the last point to the theoretical occupation $n_{\rm m}(T) \approx k_{\rm B}T/\hbar\Omega_{\rm m}$.

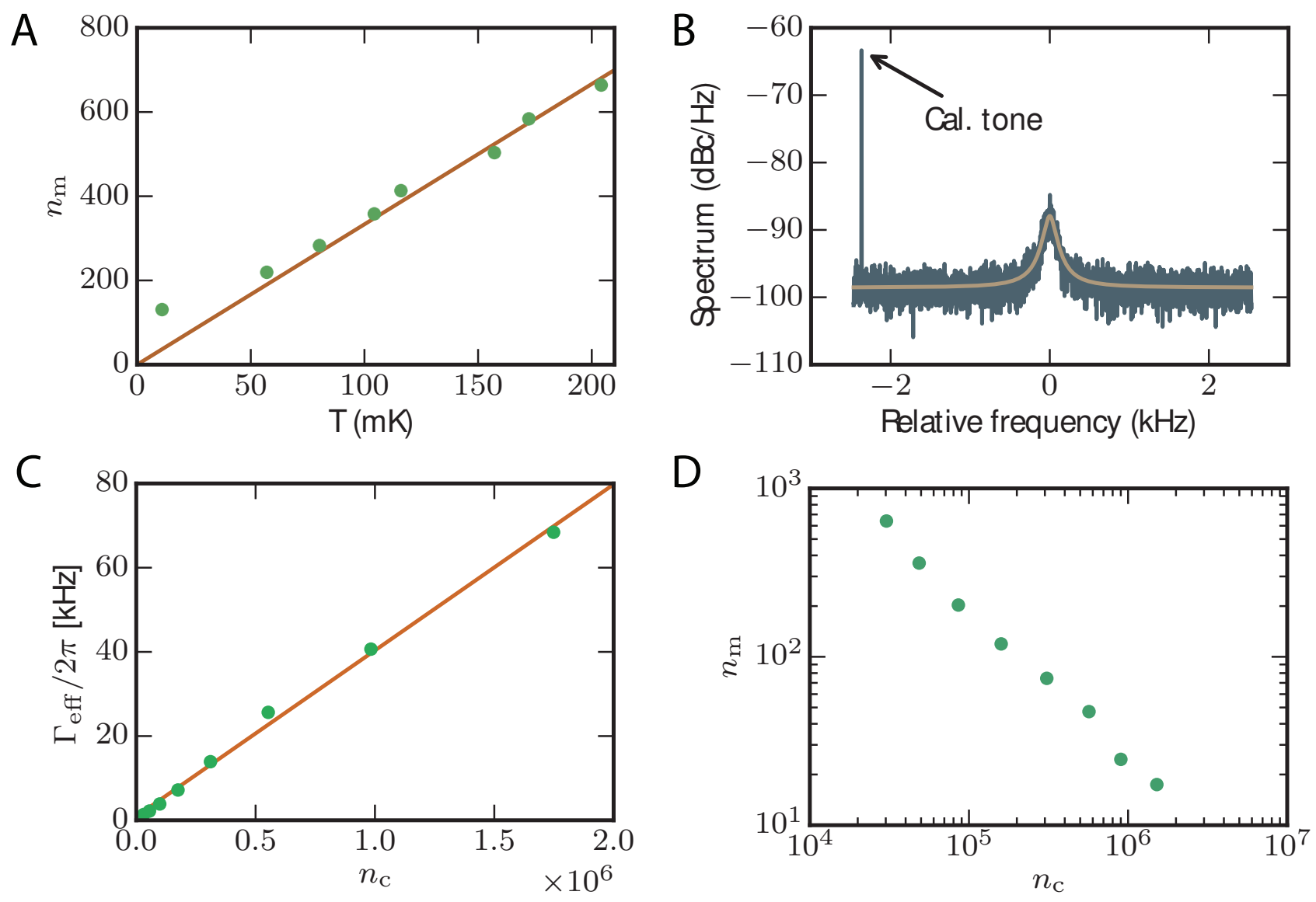

Fig. S 6. Thermalization of the mechanical oscillator and measurement of $g_0$ A. Example of thermalization of the mechanics, showing occupation of the mechanical mode as a function of refrigerator temperature. The mechanical sideband power of a resonant probe is scaled to give thermal occupation by anchoring it to the highest temperature value, where the mechanical oscillator can be assumed to be thermalized. The data is fit with a line crossing the origin representing ideal thermalization. B. Example of a mechanical sideband, with a weak tone used for calibration (cal. tone).

### B. Measurement of the vacuum coupling rate $g_0$

After the mechanical oscillator has been shown to thermalize, the same sideband measurement can be used to extract the coupling rate $g_0$. Equation (37) can be rewritten as

$$P_{\rm SB} = 4\frac{\kappa_{\rm ex}^2}{\kappa^2}\frac{1}{\Omega_{\rm m}^2+(\kappa/2)^2}g_0^2 n_{\rm m}(\beta P_{\rm in}) \tag{38}$$

where the cavity occupation $n_{\rm c}$ was expressed in terms of the probe incoming power $\beta P_{\rm in}$ at the device. The probe power $P_{\rm in}$, known at the input of the refrigerator, is attenuated along the line to the device by an imprecisely known amount $\beta$. Moreover, we measure the sideband at the output of the refrigerator, where it has be amplified by an imprecisely known gain $\mathcal{G}$. To account for $\beta$ and $\mathcal{G}$, a weak calibration tone is added close to the sideband frequency, with a known power at the input $P_{\rm cal}$, (see Fig. S6 B). The calibration tone will be amplified by the same gain $\mathcal{G}$. We assume as well that the attenuation it is subjected to is approximately the same as the probe, since the frequencies differ only by $\Omega_{\rm m}$. Therefore, by dividing the measured sideband power $P_{\rm SB}^{\rm meas} = \mathcal{G}P_{\rm SB}$ by the measured calibration tone power $P_{\rm cal}^{\rm meas} = \mathcal{G}\beta P_{\rm cal}$, the result is *independent* of both $\beta$ or $\mathcal{G}$. The vacuum coupling rate can be extracted as

$$g_0^2 = \frac{P_{\rm SB}^{\rm meas}}{P_{\rm cal}^{\rm meas}}\frac{1}{4}\left(\frac{\kappa}{\kappa_{\rm ex}}\right)^2\left(\Omega_{\rm m}^2+(\kappa/2)^2\right)\frac{1}{n_{\rm m}}\frac{P_{\rm cal}}{P_{\rm in}} \tag{39}$$

where all remaining the parameters are known or inferred. For device 2, the couplings were found to be $g_0/2\pi = 106\pm12$ Hz and $g_0^{\rm aux}/2\pi = 79 \pm 3$ Hz. The uncertainty takes into account statistical fluctuations as well as the systematic uncertainty of the thermometer.

### C. Mechanical energy decay rate ($\Gamma_{\rm m}$)

We observe a slight power dependence on the mechanical energy decay rate $\Gamma_{\rm m}$ not attributed to dynamical backaction

(i.e. even when probing on resonance of the microwave cavity). For low probe power, this dependence is linear, from which we extract the bare mechanical energy decay rate $\Gamma_{\mathrm{m}} \approx 2\pi \times 30$ Hz by extrapolation to zero for device 1 and $\Gamma_{\mathrm{m}} \approx 2\pi \times 100$ for device 2.

### D. Cooperativity ($\mathcal{C}$)

For the data in Figures 3 and 4 of the main text, we calibrate the power of the pump in terms of the cooperativity $\mathcal{C}$. We fit the the linear dependence of the energy decay rate of the microwave resonance as a function of power (on the upper sideband of the primary mode). The intercept at $\kappa_{\mathrm{eff}} = 0$ gives the threshold power $P_{\mathrm{th}}$ corresponding to $\mathcal{C} = 1$. This provides us with a self-calibrated conversion factor between input power of the pump and the cooperativity, as $\mathcal{C} = P/P_{\mathrm{th}}$.

## VIII. CALIBRATION OF THE NOISE

In this experiment, we use our HEMT amplifier as an *in situ* calibrated noise source, in order to carry out absolute noise spectra measurement. The HEMT equivalent input noise is first measured with a matched load of known temperature as its input, acting as a self-calibrated noise source. The HEMT input noise can then be used as a calibrated noise source and the noise output of our device is measured by comparison with the known background. All the noise measurements were carried out with device 2. This section details out the calibration method used.

### A. Calibration of the HEMT noise temperature

In a separate experiment, we calibrate the noise of our HEMT amplifier using a matched load as a source of Johnson noise. A 50-Ohm terminator is connected to the input of the HEMT, as shown in Fig. S7 A. We sweep the temperature of the base plate up to 10 K from base temperature. For each temperature point, the output noise power spectral density is recorded at different frequencies. An example of the result of such a sweep is presented in Fig. S7 B. The output noise is given by

$$n_{\mathrm{out}} = \mathcal{G}\mathcal{G}_{\mathrm{HEMT}} \left( n_{\mathrm{HEMT}} + n_{\mathrm{T}} \right) = \mathcal{G}\mathcal{G}_{\mathrm{HEMT}} C \left( T_{\mathrm{HEMT}} + T \right) \tag{40}$$

with $C$ a proportionality constant, $\mathcal{G}$ the total gain along the output line, and where we used the formula for the noise density of Johnson noise in order to convert all noises to effective temperatures. The noise temperature of the HEMT can then directly be compared to the actual temperature of the load $T$ by fitting the curve of Fig. S7 B. It is easily extracted as the $y$-intercept divided by the slope of a linear fit (equivalent to minus the $x$-intercept). The results for all measured frequencies are shown in Fig. S7 C and D. We compute the uncertainty on the HEMT noise combining the fitting uncertainty from statistical fluctuations and a systematic uncertainty of 16 mK from our thermometer.

### B. Output spectrum ($\bar{S}_{aa}(\omega)$)

To calibrate the output spectrum of our device $\bar{S}_{aa}(\omega)$, we use the equivalent input noise of the HEMT as a calibrated reference level. We compare the measured noise emitted by our device on resonance to the noise background, which are given by

$$N_{\mathrm{device}} = \mathcal{G}_{\mathrm{chain}} (\alpha \bar{S}_{aa}(\omega) + n_{\mathrm{HEMT}}) \tag{41}$$

$$N_{\mathrm{BG}} = \mathcal{G}_{\mathrm{chain}} n_{\mathrm{HEMT}} \tag{42}$$

where $\mathcal{G}_{\mathrm{chain}}$ is the unknown total gain from the input of the HEMT to the measurement at room temperature, $\alpha$ is attenuation on the line between the device (DUT) and the HEMT, and $n_{\mathrm{HEMT}}$ the equivalent input noise of the HEMT (see the scheme in Fig. S8 B). Since the HEMT noise dominates over other sources of noise in the chain, we can neglect them. The output spectrum can then be extracted as

$$\bar{S}_{aa}(\omega) = \frac{n_{\mathrm{HEMT}}}{\alpha} \left( \frac{N_{\mathrm{device}}}{N_{\mathrm{BG}}} - 1 \right) \tag{43}$$

in terms of the measured noises, if $n_{\mathrm{HEMT}}$ and $\alpha$ can be calibrated.

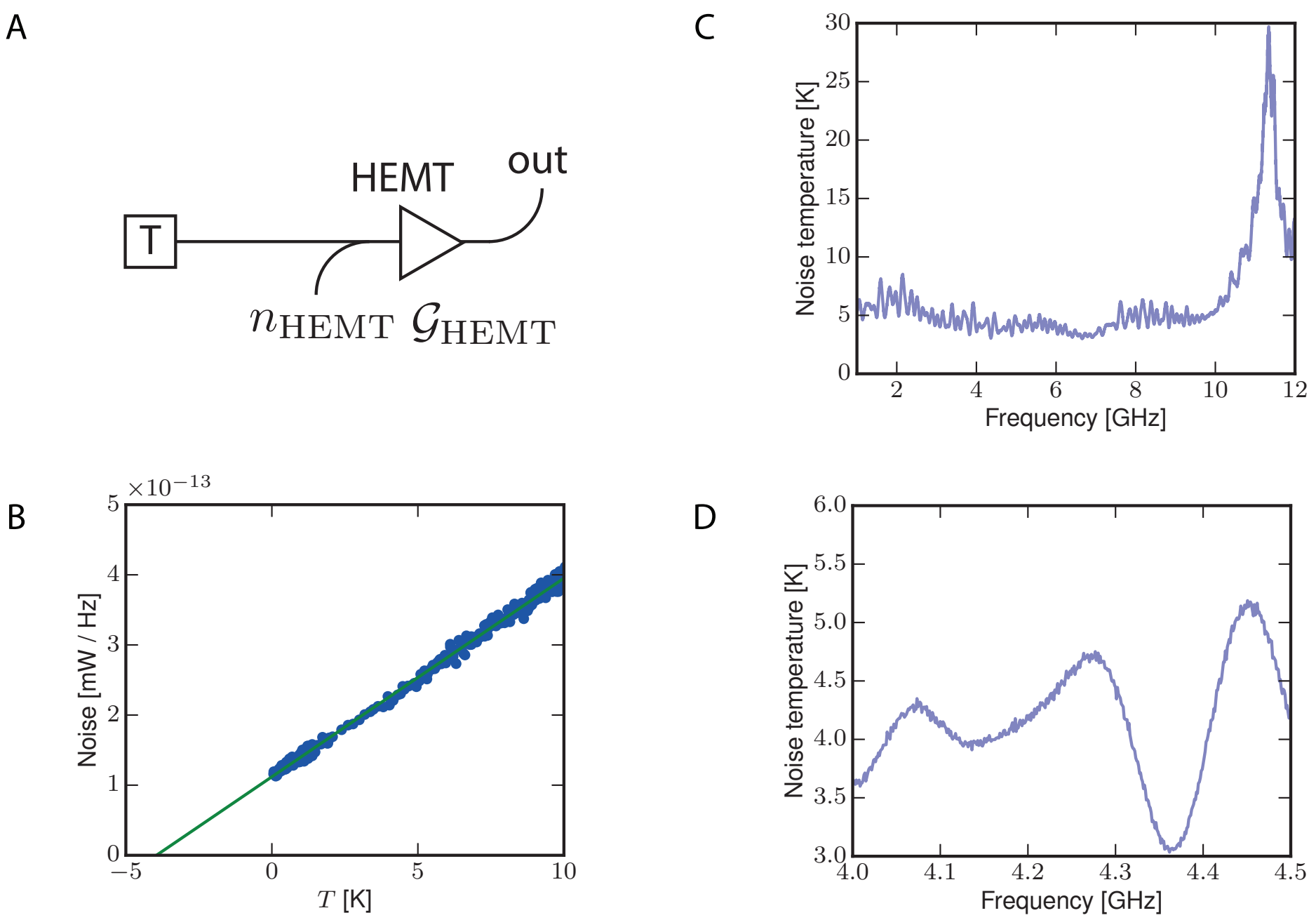

Fig. S 7. Calibration of the HEMT noise with Johnson noise. A. Scheme of noise calibration of HEMT. B. Example of temperature dependence of output noise with temperature of the resistor, for a frequency of 4.13 GHz. C. Measured HEMT noise temperature as a function of frequency. D. Zoom of C for the relevant range for our device.

Both $\alpha$ and $n_{\rm HEMT}$ are determined through independent measurements. As detailed above, the noise of the HEMT is calibrated with a temperature sweep. At 4.13 GHz, the frequency of our noise measurement, the HEMT noise temperature is $3.95 \pm 0.02$ K (see Fig. S7 D). This corresponds to $n_{\rm HEMT} = 20.0 \pm 0.1$ photons per second in a bandwidth of 1 Hz. The dominant source of uncertainty comes from the attenuation $\alpha$, as it is not possible to independently extract its value at base temperature. In a separate experiment, we measure at 77 K the attenuation of the components between the device (DUT) and the HEMT to be 1.6 dB. This is an upper bound of attenuation, as all passive components and cables are expected to show reduced attenuation at the lower base temperature. We therefore have an upper bound value for $\bar{S}_{aa}(\omega)$, providing a worst-case scenario value for the noise properties of the device.

### C. Equivalent input noise and mechanical thermal occupation

The noise performance of the amplifier is characterized by the added noise of the amplifier $\mathcal{N}$, defined as $\bar{S}_{aa}(\omega) = \mathcal{G}(\omega)(\mathcal{N}(\omega) + n_{\rm in})$. Starting with the noise output $\bar{S}_{aa}(\omega)$, we divide it by the gain curve obtained from fitting the coherent response of the device, to obtain the equivalent noise at the reference plane of the device input. In Fig. S8 C, we compare the equivalent noise at the reference plane of the HEMT input and at the input of the device. In that figure, the noise background of the HEMT is not subtracted to provide a point of reference.

### D. An independent calibration of HEMT noise using the optomechanical device

The same data, used in the previous section for characterizing thermalization of the mechanics and measuring $g_0$, can be used for absolute power calibration. We use this as an *independent* way to measure the HEMT noise background, and therefore confirm the validity of the noise calibration done above. In eq. (37), if the cavity occupation $n_{\rm c}$ is known, the power of the sideband at the device can be predicted and used as an absolute scale to measure the noise background. This provides an independent measure of HEMT noise divided by the attenuation $\frac{n_{\rm HEMT}}{\alpha}$: the equivalent

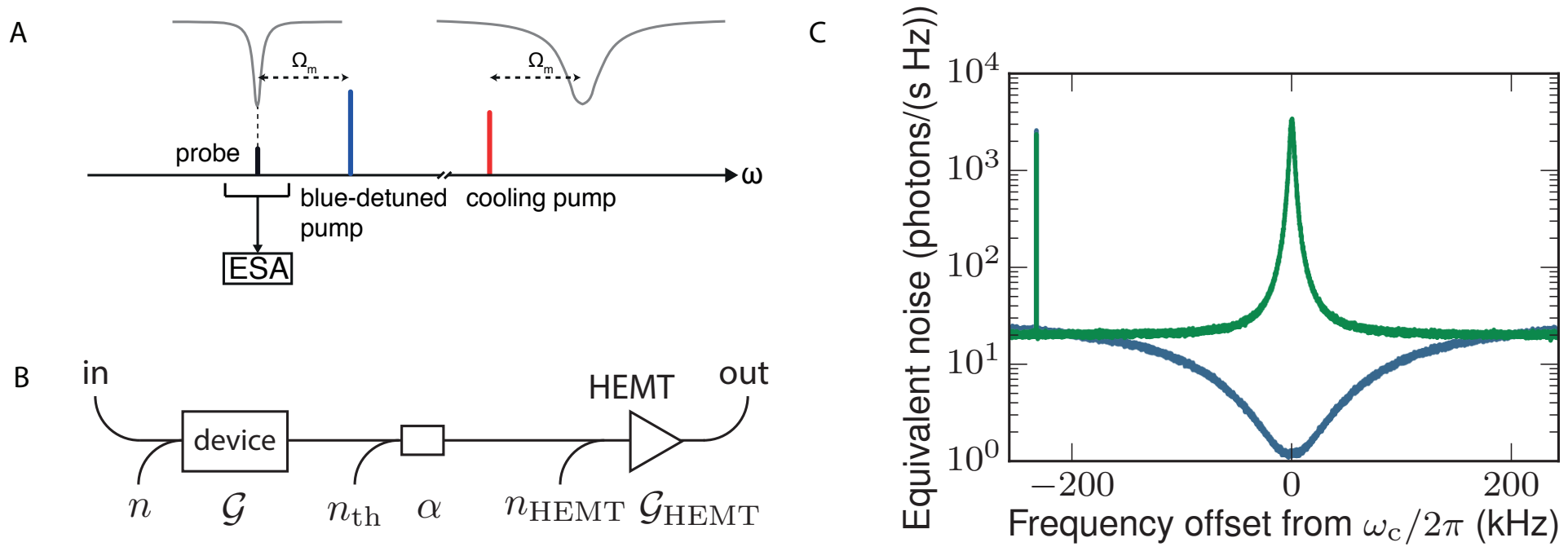

Fig. S 8. Calibration of added noise in amplification. A. Frequency scheme of the measurement. The noise spectrum emission of the cavity is measured while a weak probe is scanned across the resonance. B. Schematics of the amplification chain, with our device connected in series with the HEMT, including attenuation. C. Output noise at a gain of 36 dB renormalized to represent equivalent noise at the input of the HEMT (green) and equivalent noise at the input of the device (blue). Here we do not subtract the HEMT noise, so that all system noise is represented and not only added noise.

noise of the HEMT at the device output.

To measure the cavity occupation $n_c$, a secondary measurement is performed and OMIT is measured with a probe of the same power now on the red sideband of the cavity resonance. The width of the mechanics is fitted to obtain

$$\Gamma_{\text{eff}} = \Gamma_m + 4\frac{g_0^2 n_{c0}}{\kappa} = \Gamma_m + 4\frac{g_0^2}{\kappa} n_c \frac{(\kappa/2)^2}{\Omega_m^2 + (\kappa/2)^2} \tag{44}$$

where $n_c$ is the cavity occupation for the resonant probe and $n_{c0}$ is the cavity occupation for the red-detuned probe used for OMIT. With this inferred cavity occupation $n_c$, the absolute rate of scattered photons of the sideband can be predicted as

$$\frac{P_{\text{SB}}}{\hbar\omega} = \frac{\kappa_{\text{ex}}}{\Omega_m^2 + (\kappa/2)^2} g_0^2 n_c n_m \tag{45}$$

in units of photons per second. By rescaling the measured sideband to the predicted power at the device output, the background is scaled to the equivalent noise at the device input. We perform this measurement with the main mode, at temperatures above thermalization point, we obtain $\frac{n_{\text{HEMT}}}{\alpha} = 42 \pm 4$ photons / (s Hz), with the uncertainty given by the statistical fluctuations of measurements. We use a value of $31.5 \pm 0.2$ photons / (s Hz), obtained from the measured HEMT noise temperature of $4.27 \pm 0.03$ K (at 4.08 GHz frequency) and 1.6 dB of estimated attenuation. This value is different than for the measurement in the main text (3.95 K at 4.13 GHz), as the cavity resonance had drifted between the two measurements. The discrepancy of 1.4 dB between the two calibration methods is attributed to systematics for this measurement. In particular, it is assumed that the probe on resonance and on the red sideband of the resonance is attenuated the same way in order to infer $n_c$, which could yield a systematic error. We stress that this represents a *fully independent* check for the validity of the calibration procedure used for noise in this experiment.

### E. Thermal occupation of the primary microwave cavity

With strong pumps near the auxiliary cavity to damp the mechanical oscillator, heating of the primary microwave mode is observed. The dual mode structure of the device enables to characterize this noise using an out-of-loop measurement. With a cooling pump at the auxiliary mode and no pump present at the primary mode, noise emission from the primary cavity is measured. Subtracting the background of HEMT noise (measured with all pumps turned off), the emission from the cavity is shown in Fig. S9 A. A Lorentzian peak of noise is emitted from the cavity, increasing in size with pump power. We interpret this as the thermal bath of the cavity mode ($n_{\text{cav}}$) effectively being heated by the pump. The excess total thermal noise measured corresponds to $\frac{\kappa_{\text{ex}}\kappa_0}{\kappa} n_{\text{cav}}$ in photons per second. Fitting the Lorentzian peak to extract the total emitted photon rate, we infer the cavity thermal occupation due to spurious heating. For a power corresponding to $\Gamma_{\text{eff}}/2\pi \approx 400$ kHz, as used for the measurements with device 2, the heating corresponds to $n_{\text{cav}} \approx 1$ of thermal microwave occupation, i.e. a noise temperature of 200 mK. We note that this should be considered as a lower bound on $n_{\rm cav}$ when the device is operated as an amplifier, as the additional tone around the primary mode can induce more heating. This implies that the number for the effective thermal occupation of the reservoir, $n_{\rm eff} = 0.41$ (cf. main text), is in fact an upper bound.

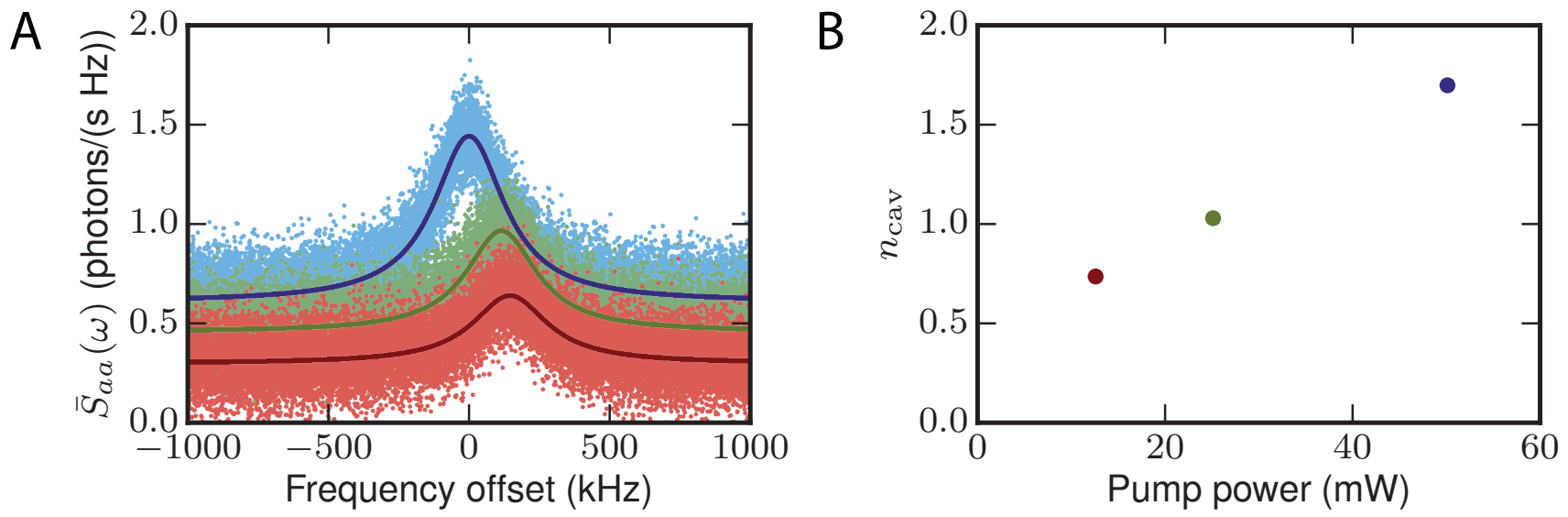

Fig. S 9. Cavity thermal occupation A. Thermal noise emission from the primary cavity while a pump is a applied to the red sideband of the auxiliary cavity. Increasing pump powers are shown from red to blue. The middle power (green) corresponds to $\Gamma_{\rm eff}/2\pi \approx 400$ kHz used for the experiment in the reversed dissipation regime. B. Inferred cavity thermal occupation from fitting the Lorentzian peak, for the same pump powers.

## IX. IMPACT OF NOISE ON THE MICROWAVE SOURCE

We compute here the limits of cooling for the final phonon occupancy due to the phase noise of the microwave sources reaching the auxiliary, cooling mode. We note that the noise reaching the primary mode is already taken into account in our analysis through $n_{\rm in}$, and does not constitute added noise in the amplification process.

We start from the Langevin equations describing the interaction between the auxiliary mode and the mechanical oscillator

$$\dot{\delta\hat{a}} = +i\Delta\delta\hat{a} - \frac{\kappa}{2}\delta\hat{a} + ig\delta\hat{b} + \sqrt{\kappa_0}\,\hat{a}_{\rm in,ex} + \sqrt{\kappa_{\rm ex}}\,\hat{a}_{\rm in,0} \tag{46}$$

$$\dot{\delta\hat{b}} = -i\Omega_{\rm m}\delta\hat{b} - \frac{\Gamma_{\rm m}}{2}\delta\hat{b} + ig\delta\hat{a} + \sqrt{\Gamma_{\rm m}}\,\hat{b}_{\rm in} \tag{47}$$

where $\Delta = -\Omega_{\rm m}$, $g = g_0\sqrt{n_{\rm c}}$, and $\delta\hat{a}$ is now the annihilation operator for the auxiliary mode. For simplicity of notation, we write in this section $\kappa$, $\kappa_0$ and $\kappa_{\rm ex}$ to refer to the auxiliary mode. Taking the Fourier transform and solving for $\delta\hat{b}$, one obtains

$$\delta\hat{b}(\omega) = \frac{1}{i(\Omega_{\rm m}-\omega)+(\Gamma_{\rm m}+\Gamma_{\rm om})/2}\left(\sqrt{\Gamma_{\rm m}}\hat{b}_{\rm in}(\omega) + i\sqrt{\Gamma_{\rm om}}\sqrt{\frac{\kappa_0}{\kappa}}\hat{a}_{\rm in,ex}(\omega) + i\sqrt{\Gamma_{\rm om}}\sqrt{\frac{\kappa_{\rm ex}}{\kappa}}\hat{a}_{\rm in,0}(\omega)\right) \tag{48}$$

where the optomechanical coupling rate of the mechanics is given by $\Gamma_{\rm om} = 4g^2/\kappa$. The noise input of the auxiliary mode is transduced to the mechanics and should be taken into account as an effective mechanical noise, when considering the interaction of the mechanical oscillator with the primary mode in eq. (15). In terms of equivalent noise, this translates to

$$n_{\rm eq} = \frac{\Gamma_{\rm m}n_{\rm m} + \Gamma_{\rm om}\frac{\kappa_0}{\kappa}n_{\rm in} + \Gamma_{\rm om}\frac{\kappa_{\rm ex}}{\kappa}n_{\rm cav}}{\Gamma_{\rm m}+\Gamma_{\rm om}} \tag{49}$$

where $n_{\rm in}$ and $n_{\rm cav}$ refer to the auxiliary mode. The microwave noise is part of the effective bath that the mechanical oscillator equilibrates to, at the same time as the cavity damps the intrinsic mechanical noise. The added noise due to microwave noise reaching the auxiliary cavity amounts to

$$n_{\rm added} = 4\frac{\kappa_0}{\kappa}\frac{g_0^2 n_{\rm c}}{\kappa\Gamma_{\rm eff}}n_{\rm in} \tag{50}$$

with $\Gamma_{\rm eff} = \Gamma_{\rm m} + \Gamma_{\rm om}$. The cavity occupation due to the red-detuned pump $n_{\rm c}$ and the microwave noise $n_{\rm in}$ can both be expressed in terms of the pump flux of photons $P/\hbar\omega$ as

$$n_{\rm c} = \frac{\kappa_{\rm ex}}{(\kappa/2)^2+\Omega_{\rm m}^2}\frac{P}{\hbar\omega} \quad \text{and} \quad n_{\rm in} = \frac{P}{\hbar\omega}\mathcal{S}_{\varphi\varphi} \tag{51}$$

where $\mathcal{S}_{\varphi\varphi}$ the phase noise of the source expressed in power below the carrier in a unit bandwidth (dBc/Hz). The added phonons from phase fluctuations after cooling amount therefore to

$$n_{\mathrm{added}} = 4\left(\frac{\kappa_{\mathrm{ex}}}{\kappa}\right)^2 \frac{g_0^2}{\Omega_{\mathrm{m}}^2 + (\kappa/2)^2}\left(\frac{P}{\hbar\omega}\right)^2 \frac{1}{\Gamma_{\mathrm{eff}}}\mathcal{S}_{\varphi\varphi}. \tag{52}$$

The power of the pump reaching each resonator $P$ is not an independent variable, but constrained in our experiment. For the cooling pump at the auxiliary mode, the pump damps mechanical mode at a rate $\Gamma_{\mathrm{eff}} \approx \Gamma_{\mathrm{om}}$ of $2\pi \times 400$ kHz in the experiment. The required power for this cooling rate is then given by

$$\frac{P}{\hbar\omega} = \frac{1}{4}\frac{(\kappa/2)^2 + \Omega_{\mathrm{m}}^2}{g_0^2}\frac{\kappa}{\kappa_{\mathrm{ex}}}\Gamma_{\mathrm{eff}}. \tag{53}$$

Inserting the expression in eq. (52), we obtain [7]

$$n_{\mathrm{added}} = \frac{1}{4}\frac{(\kappa/2)^2 + \Omega_{\mathrm{m}}^2}{g_0^2}\Gamma_{\mathrm{eff}}\mathcal{S}_{\varphi\varphi}. \tag{54}$$

The maximum allowed phase noise to increase the occupation by less than 1 is

$$\mathcal{S}_{\varphi\varphi}^{\mathrm{max}} = 4\frac{1}{(\kappa/2)^2 + \Omega_{\mathrm{m}}^2}\frac{g_0^2}{\Gamma_{\mathrm{eff}}}. \tag{55}$$

For the microwave source used, the phase noise was measured to be -147 dBc/Hz at a detuning of 6 MHz. With the parameters of our system, this amounts to added noise of $n_{\mathrm{added}} \approx 9$. This justifies the necessity of filter cavities to reduce the noise within the bandwidth of the cavity by more than 10 dB in order to reach occupations below 1.

---